\documentclass{aa}

\usepackage[utf8]{inputenc}
\usepackage[varg]{txfonts}
\usepackage{xcolor}
\usepackage{lscape}
\usepackage{pgffor}
\usepackage{array}
\usepackage{subcaption}
\usepackage{tablefootnote}

\graphicspath{{Images/}}

\newcommand{\software}[1]{\texttt{#1}}
\newcommand{\size}{0.95}

\date{\today}

\begin{document}

    \title{{The AMIGA sample of isolated galaxies }}
    \subtitle{XIV. Disc breaks and interactions through ultra-deep optical imaging} 
    
    \titlerunning{Disc breaks and interactions through ultra-deep optical imaging}
    
    \author{P.~M.~Sánchez-Alarcón \inst{1,2}\fnmsep \thanks{\email{pmsa.astro@gmail.com}}
          \and
          J.~Román \inst{1,2,3,4}\fnmsep \thanks{\email{jromanastro@gmail.com}}
          \and 
          J.~H.~Knapen \inst{1,2}          
          \and
          L.~Verdes-Montenegro \inst{3}
          \and
          S.~Comerón \inst{2,1}
          \and
          R.~M.~Rich \inst{5}
          \and
          J.~E.~Beckman \inst{1,2}
          \and 
          M.~Argudo-Fernández \inst{6,7}
          \and 
          P.~Ramírez-Moreta \inst{8,9}
          \and 
           J.~Blasco \inst{6}
           \and 
           E. Unda-Sanzana \inst{10}
           \and
           J.~Garrido \inst{3}
           \and 
          S.~Sánchez-Exposito \inst{3}.
          }
    \authorrunning{Sanchez-Alarcon, PM et al.}

   \institute{Instituto de Astrofísica de Canarias, c/ Vía Láctea s/n, E-38205, La Laguna, Tenerife, Spain
             \and
             Departamento de Astrofísica, Universidad de La Laguna, E-38206, La Laguna, Tenerife, Spain
             \and
             Instituto de Astrofísica de Andalucía (CSIC), Granada, Spain
             \and 
             Kapteyn Astronomical Institute, University of Groningen, PO Box 800, NL-9700 AV Groningen, the Netherlands
             \and
             Department of Physics \& Astronomy, University of California Los Angeles, 430 Portola Plaza, Los Angeles, CA 90095-1547, USA
             \and
             Departamento de Física Teórica y del Cosmos Universidad de Granada, 18071 Granada, Spain
            \and
            Instituto Universitario Carlos I de F\'isica Te\'orica y Computacional, Universidad de Granada, 18071 Granada, Spain 
             \and
             ESA NEO Coordination Centre, Via Galileo Galilei, 00044 Frascati (RM), Italy
             \and
             GMV, Isaac Newton 11, Tres Cantos, 28760 Madrid, Spain
             \and
             Centro de Astronomía (CITEVA), Universidad de Antofagasta, Avda. U. de Antofagasta 02800, Antofagasta, Chile
             }

   \date{\today }

 
  \abstract{In the standard cosmological model of galaxy evolution, mergers and interactions play a fundamental role in shaping galaxies. Galaxies that are currently isolated are thus interesting, allowing us to distinguish between internal and external processes affecting the galactic structure. However, current observational limits may obscure crucial information in the low-mass or low-brightness regime.}
  {We use optical imaging of a subsample of the AMIGA catalogue of isolated galaxies to explore the impact of different factors on the structure of these galaxies. In particular, we study the type of disc break as a function of the degree of isolation and the presence of interaction indicators like tidal streams or plumes which are only detectable in the ultra-low surface brightness regime.}
  {We present ultra-deep optical imaging in the \textit{r}-band of a sample of 25 low-redshift ($z < 0.035$) isolated galaxies. Through careful data processing and analysis techniques, the nominal surface brightness limits achieved are comparable to those to be obtained on the 10-year LSST coadds ($\mu_{r,\textrm{lim}}$~$\gtrsim$~29.5~$\textrm{mag~arcsec}^{-2}$ [3$\sigma$; 10"$\times$10"]). We place special emphasis on preserving the low surface brightness features throughout the processing. }
  {The extreme depth of our imaging allows us to study the interaction signatures of 20 galaxies, given that the presence of Galactic cirrus is a strong limiting factor in the characterisation of interactions for the remaining 5 of them. We detect previously unreported interaction features in 8 (40\%$\pm$14\%) galaxies in our sample.
  We identify 9 galaxies (36\%$\pm$10\%) showing an exponential disc (Type~I), 14 galaxies (56\%$\pm$10\%) with down-bending (Type~II) profile and only 2 galaxies (8\%$\pm$5\%) with up-bending (Type~III) profiles. 
  Isolated galaxies have considerably more purely exponential discs and fewer up-bending surface brightness profiles than field or cluster galaxies. We find clear minor merger activity in some of the galaxies with single exponential or down-bending profiles, and both of the galaxies with up-bending profiles show signatures of a past interaction.}
  {We show the importance of ultra-deep optical imaging in revealing the presence of faint external features in galaxies which indicate a probable history of interaction. We confirm that up-bending profiles are likely produced by major mergers while down-bending profiles are probably formed by a threshold in star formation. Unperturbed galaxies, evolving slowly with a low star formation rate could induce the high rate of Type~I discs in isolated galaxies.}


\maketitle

\section{Introduction}

Present-day galactic discs are snapshots of galaxy evolution resulting from galaxy formation and evolution through cosmic time. The study of galaxies with a variety of morphologies, evolutionary stages, and environments helps us to assemble a global picture to explain all the observed features. One of the simplest, yet most effective ways to classify galactic structures is through their surface brightness profiles. These profiles were initially characterised with a single exponential decay by \cite{1940BHarO.914....9P} and \cite{1958ApJ...128..465D}, but the current view is considerably more complex. Following the observation of breaks in profiles by \cite{Kruit79} and \cite{Kruit81,Kruit81b} at the outer regions of the edge-on galaxies, work by \cite{Erwin05} and \cite{Pohlen06} considered breaks over a range of galactocentric radii and in less-inclined galaxies. They established a classification based on the global shape of the disk profile: Type~I or pure exponential for profiles with no breaks, Type~II or down-bending profiles with outer slopes steeper than the inner ones \citep[e.g.][]{Freeman70, Pohlen02} and Type~III or up-bending profiles where the exponential decline is steeper in the inner part of the disc that in the outer part \citep[e.g.][]{Erwin05}. This classification has been helpful since galaxies show different average properties for each type, suggesting a different origin or evolution.

One of the most characteristic features in discs is the drastic change in age around the break radius, noticed as a "U-shape" in the colour profiles, while the mass surface density profile remains relatively constant \citep[e.g.][]{Azzollini08, Bakos08, Bakos12, Martin-Navarro12, Zheng15, Watkins16, 2016MNRAS.456L..35R}. This observational feature has been tentatively explained as the result of gas accretion plus a density threshold in the star formation, and subsequent redistribution of mass by radial migration \citep[e.g][]{Debattista06,Roskar08, MS09} which has been described as inside-out formation of the disc \citep[][]{SB09}. 

The environment appears to play a key role in the frequency of each break type \citep[e.g.][]{Pranger17, Watkins19}. In general, high-density environments favour up-bending Type~III profiles \citep[see][]{Maltby12}. This is not surprising, given that the environment is one of the most influential factors in shaping galactic morphology \citep[e.g., ][]{2004ApJ...615L.101B, 2005ApJ...621..673T, 2006MNRAS.373..469B, Erwin12}. However, internal processes \citep[e.g., ][]{2004ARA&A..42..603K} or the accretion of cold gas \citep[][]{2017ASSL..434..209B} are also capable of transforming the characteristics of the galaxies.  Correlations between the type of break (in particular the Type~III fraction) with internal parameters \citep{2017MNRAS.470.4941H, 2018MNRAS.479.4292W}, and possible interactions \citep{Eliche-Moral15, Borlaff18} are also found. Adding more complexity, instrumental effects including background subtraction or scattered light can have a significant effect on the photometric profiles, especially at extremely low surface brightness \citep[][]{2014A&A...567A..97S}. This means that determining the specific impact of different processes is not a straightforward task. A possible approach is to study isolated galaxies, excluding or diminishing the impact of environmental processes \citep[e.g., ][]{1973SoSAO...8....3K, 1977ApJ...216..694H, 1981Afz....17...53A, 2001A&A...373..402S, 2004A&A...420..873V}. 

Determining and quantifying the isolation of a galaxy is not a simple task. The Analysis of the interstellar Medium of Isolated GAlaxies (AMIGA) Project \citep[][]{2005A&A...436..443V} is an exhaustive study of galaxies isolated from major companions, based on the original catalogue of isolated galaxies (CIG) presented by \cite{1973SoSAO...8....3K} and later revised and quantified by \cite{2007A&A...470..505V, Verley-Iso} and \cite{Argudo-Fernandez13}. Despite the efforts shown in these works to quantify minor interaction features, low signal-to-noise spectroscopic and imaging surveys may fail to identify the presence of faint surface brightness satellites around galaxies classified as isolated. In fact, numerical simulations based on the $\Lambda$-CDM cosmological paradigm predict an average of one low surface brightness feature per galaxy due to minor interactions at a surface brightness level of $\mu$ = 29\,$\textrm{mag\,arcsec}^{-2}$ \citep[][]{2001ApJ...557..137J} and we can expect many galaxy features fainter than $\mu = 30\,\textrm{mag\,arcsec}^{-2}$ \citep[][]{2008ApJ...689..936J}. 

The interest in low surface brightness science is confronted with significant challenges due to numerous observational limitations. Advances have been made in recent years in developing deep optical surveys \citep[e.g., ][]{Laine14, Dragonfly, 2016MNRAS.456.1359F, 2016MNRAS.460.1270D, 2018RNAAS...2..144R, HSC-SSP, Martinez-Delgado19, 2021A&A...654A..40T}, characterising scattered light \citep[e.g., ][]{2009PASP..121.1267S, 2017A&A...601A..86K, 2020MNRAS.491.5317I}, improving observational \citep[][]{2010A&A...513A..78J, 2015MNRAS.446..120D, 2017ApJ...834...16M, Trujillo16,2019MNRAS.486.1995S} and data processing techniques \citep{2014ApJS..213...12J, Trujillo16, Borlaff19, 2021A&A...645A.107H}, and more \citep[see review by][]{2019arXiv190909456M}. Indeed, increasingly sophisticated studies are able to reveal the presence of satellites \citep[][]{Laine14, 2015ApJ...809L..21M, 2021A&A...656A..44R} and tidal features \citep[][]{Laine14, Martinez-Delgado19, 2023A&A...671A.141M, 2022ApJS..262...39H, 2023arXiv230503073R} at increasingly lower surface brightness, but a long way remains to correctly detect and classify all the structures predicted by simulations \citep[see][and references therein]{2022MNRAS.513.1459M}.

In this work, we carry out an ultra-deep imaging study of isolated galaxies from the AMIGA catalogue in order to shed light on the structural differences between these isolated galaxies and galaxies in higher-density environments. In particular, we are interested in: 1) the types of breaks in the discs of isolated galaxies in comparison with those in higher-density environments. This is something not yet done in the literature for isolated galaxies, and so far has only been carried out comparing higher-density environments such as clusters, and groups with simply the "field". 2) Using low surface brightness imaging to explore the presence of minor interactions in these isolated galaxies. 3) Explore possible correlations between the type of break, presence of minor interactions and density, that can help elucidate the dominant factors in shaping the disc of galaxies.

This work is structured as follows: In Section~\ref{sec:Sample} we describe the data sample and the reduction procedure followed. In Section~\ref{sec:Analysis} we explain the methods used to measure the surface brightness profiles, classify the profiles by type and detect signs of interaction. In Section~\ref{sec:Results} we present the results that are discussed in Section~\ref{sec:Discussion}. The conclusions of our work are summarised in Section~\ref{sec:Conclusions}. We adopt the values of the cosmological constants $H_0 = 70\,\textrm{km\,s}^{-1}\,\textrm{Mpc}^{-1}$, $T_{0}=2.725\,\textrm{K}$, and $\Omega_{m}=0.3$. Galactic extinction is corrected following \citet[][]{2011ApJ...737..103S}. We use the AB photometric system.

\section{Sample, data and processing}
\label{sec:Sample}

\begin{table*}[h!]
\centering
\caption{Observational and physical properties of the 25 galaxies studied in this work.}
\begin{tabular}{llcccccccc}

\hline \hline \\[-8pt]
Galaxy & ID & $V_{\textrm{rad}}$ & $L_B$   & Telescope/Survey & Integration time & Depth $\mu_{\textrm{lim}}$ & FOV  \\
 &  & [km\,s$^{-1}$] & [log ${\rm L}_\odot$]  &  & [hours] & [mag\,arcsec$^{-2}$ ] & degrees$^2$ \\
\hline \hline\\[-8pt]
CIG 11 & UGC 139 & 3906 & 10.03 & VST & 4.8 & 29.7 & 4.936 & \\
CIG 33 & NGC 237 & 4090 & 9.95 & VST & 4.8 & 29.7 & 4.987 & \\
CIG 59 & UGC 1167 & 4193 & 9.56 & VST & 4.8 & 29.7 & 6.040 & \\
CIG 94 & UGC 1706 & 4694 & 9.74 & INT & 4.8 & 29.8 & 0.451 & \\
CIG 96 & NGC 864 & 1562 & 9.57 & VST & 5.9 & 29.7 & 6.041 & \\
CIG 100 & UGC 1863 & 6546 & 9.86 & HSC-SSP & - & 29.5 & 0.155 & \\
CIG 154 & UGC 3171 & 4530 & 10.06 & VST & 4.8 & 29.5 & 6.006 & \\
CIG 279 & NGC 2644 & 2196 & 9.03 & VST & 4.8 & 29.5 & 6.046 & \\
CIG 329 & NGC 2862 & 4473 & 9.27 & INT & 3.5 & 29.8 & 0.318 & \\
CIG 335 & NGC 2870 & 3568 & 10.17 & INT & 6.2 & 30.0 & 0.340 & \\
CIG 340 & IC 2487 & 4710 & 9.61 & INT & 5.7 & 30.3 & 0.418 & \\
CIG 512 & UGC 6903 & 2234 & 8.65 & HSC-SSP & - & 29.6 & 0.155 & \\
CIG 568 & UGC 8170 & 11094 & 10.02 & VST & 6.8 & 30.4 & 0.598 & \\
CIG 613 & UGC 9048 & 11121 & 10.72 & INT & 4.1 & 30.0 & 0.307 & \\
CIG 616 & UGC 9088 & 6668 & 9.65 & INT & 3.8 & 30.0 & 0.308 & \\
CIG 626 & NGC 5584 & 1934 & 9.43 & HSC-SSP & - & 29.8 & 0.155 & \\
CIG 744 & UGC 10437 & 3088 & 9.16 & HSC-SSP & - & 29.6 & 0.155 & \\
CIG 772 & IC 1231 & 5643 & 10.34 & INT & 4.7 & 30.1 & 0.309 & \\
CIG 800 & NGC 6347 & 6636 & 10.08 & INT & 4.2 & 29.9 & 0.307 & \\
CIG 838 & IC 1269 & 6558 & 10.74 & INT & 3.6 & 29.6 & 0.315 & \\
CIG 947 & NGC 7217 & 952 & 10.52 & JRT & 10.3 & 29.0 & 0.142 & \\
CIG 971 & UGC 12082 & 802 & 8.48 & JRT & 3.0 & 28.5 & 0.118 & \\
CIG 1002 & NGC 7451 & 6638 & 9.96 & VST & 10.0 & 29.9 & 7.297 & \\
CIG 1004 & NGC 7479 & 2443 & 9.70 & JRT & 3.7 & 28.5 & 0.127 & \\
CIG 1047 & UGC 12857 & 2465 & 9.45 & VST & 6.2 & 29.6 & 6.456 & \\

\hline 
\end{tabular}
\tablefoot{The first two columns show the CIG name and the main ID of each galaxy. The third and fourth columns show the radial velocity and the absolute luminosity in the \textit{B}~band of the galaxy, obtained from the last revision of the AMIGA parameters \cite{2018A&A...609A..17J}. The fifth column shows the telescope used to acquire the images (see Section~\ref{sec:Observations}). The sixth and seventh columns show the total integration time in hours and the surface brightness limits [$3\sigma,\,10^{\prime\prime}\times10^{\prime\prime}$] of the image in $\textrm{mag\,arcsec}^{-2}$ computed as explained in \cite{Roman20}, respectively. The eighth column shows the field of view (FOV) of the whole image in degrees$^2$.}
\label{tab:data_sample}
\end{table*} 

\subsection{Sample selection and observations}
\label{sec:Observations}

We use the AMIGA sample \citep[][]{2005A&A...436..443V}, based on the original CIG catalogue by \cite{1973SoSAO...8....3K}. The latest revision of the isolation parameters of the AMIGA catalogue was carried out by \cite{Argudo-Fernandez13}, a work on which we base the selection of our sample.

Our selection criteria are as follows. The targets belong to the AMIGA catalogue. We require that the galaxies have a reliable determination of the distance \citep[latest revision for AMIGA sample in][]{2018A&A...609A..17J} as well as a detection in H\,{\sc i}, in order to also further explore possible correlations between the fainter optical morphology and their gas content. The photometric isolation parameters measured by \cite{Verley-Iso} are the local number density of neighbours galaxies, $\eta_{k,\textrm{p}}$ and the tidal strength, $Q_{\textrm{Kar,p}}$.  We require galaxies to have a local number density of neighbouring galaxies $\eta_{k,\textrm{p}}<2.7$.

We observed 25 galaxies meeting these criteria with different telescopes. These galaxies have morphologies that range from Hubble types of $3\leq\rm{T}\leq5$, which is a similar sample of the complete AMIGA catalogue \citep[see][]{2006A&A...449..937S}.  Here we briefly describe the instrumentation used: 1) The Isaac Newton Telescope (INT), located in the Observatorio del Roque de los Muchachos in La Palma, Spain, has a 2.5\,m diameter primary mirror. We used the Wide Field Camera (WFC), a four-CCD mosaic covering 33~arcmin on a side with a pixel scale of 0$\farcs$333. 2) The VLT Survey Telescope (VST), located in Cerro Paranal, Chile. The VST has a 2.65\,m primary mirror. We used OmegaCAM which has 32 CCDs covering a field-of-view of approximately 1 degree$^2$, with a pixel scale of 0$\farcs$21. 3) The 0.7\,m Jeanne-Rich Telescope (JRT) is located at the Polaris Observatory Association site, Pine Mountain, California. The camera covers approximately 40\,arcmin on a side with a pixel scale of 1$\farcs$114.

Due to the high observational cost of ultra-deep optical imaging, and to optimise the detection of faint features we only obtained \textit{r}~band data. Observations were carried out mostly in dark time, although grey nights with little Moon were also used. In the worst cases, the Moon was far enough with a low phase to not produce any gradient in the image. Dithering patterns of tens of arcseconds were used in order to improve the flat field. The exposure time was set between one and five minutes for each individual frame. 

The imposed surface brightness limits rule out the use of most general-purpose optical surveys. Only the Hyper Suprime-Cam Subaru Strategic Program (HSC-SSP) \citep{HSC-SSP} is able to fulfil our surface brightness requirements. HSC-SSP is a survey produced with the Subaru 8.2\,m aperture telescope and the Hyper Suprime-Cam. We found four galaxies within the HSC-SSP footprint meeting our selection criteria, that are included in our sample.  We used the second data release of the survey \citep[][]{HSC-SSP-2}. Although there are additional filters, for the purposes of our work we only use the \textit{r} band data.

We set a requirement of $\mu_{r,\textrm{lim}} > 29.5\,\textrm{mag\,arcsec}^{-2}$ for the surface brightness  measured as 3$\sigma$ in $10^{\prime\prime}\times10^{\prime\prime}$ boxes following the nominal depth description by \cite{Roman20}. This limit is set to have a compromise between deep images to detect faint structures and observational costs. We add an exception to this limit to data from the JRT. This telescope has a small aperture but is built to be extremely efficient in the low surface brightness regime. Therefore, although the Poissonian noise with which the surface brightness limits are measured will tend to be higher for this telescope, the detectability of extremely low surface brightness features is comparable to data of higher nominal depth, as we will show in the results section. Additionally, on this telescope, we use a broad luminance band, in order to maximise detection. This type of band has been proven to be similar to \textit{r}-band \cite{2015AJ....150..116M}.  We therefore set a limit of $\mu_{L,\textrm{lim}} > 28.5\,\textrm{mag\,arcsec}^{-2}$ for data from the JRT.

\begin{figure*}[ht!] 
	\includegraphics[width=\linewidth]{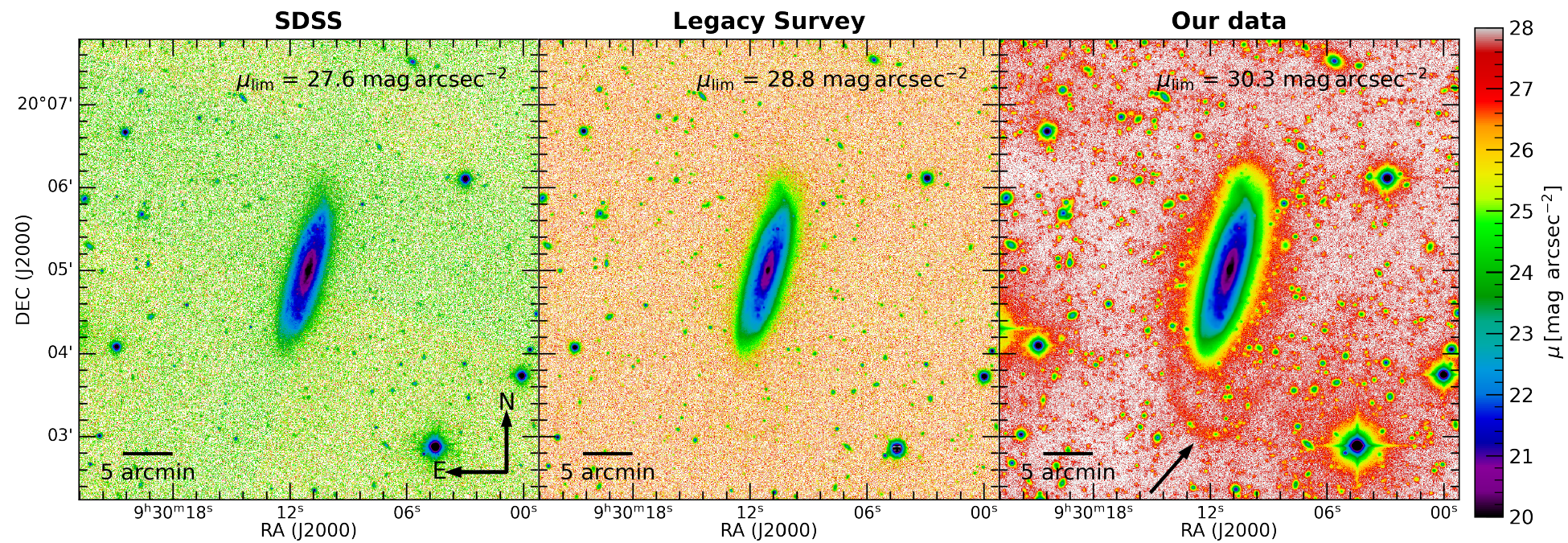}
	\caption{Comparison between the Sloan Digital Sky Survey (left), the Legacy Survey (middle) and our data for the galaxy CIG~340 (IC2487). The upper-right value indicates the surface brightness limit [$3\sigma, 10^{\prime\prime}\times10^{\prime\prime}$] of each image computed as explained in Section~\ref{sec:Sample}. The arrow in the left pannel indicates the low surface brightness feature only visible in our data.}
	\label{fig:Comparison}
\end{figure*}
\subsection{Data Reduction}\label{sec:data_reduction}
The observations were reduced following a procedure aimed at preserving the low surface brightness features. First, a bias subtraction (and dark if necessary) is performed, following an ordinary procedure of combining bias images (and dark). For the flat-fielding, we use the science images themselves. This procedure consists of using heavily masked science images, masked with specialised software such as \software{SExtractor} \citep[][version 2.25.0]{Bertin1996} and \software{Noisechisel} \citep[][version  0.18]{gnuastro}, which are normalised in flux and subsequently combined using a resistant mean algorithm to produce a flat that is representative of the sky background during the observations. This procedure, building the flat from the science images, has significant advantages over the usual dome or twilight flats. The main one is a considerable decrease in the strength of gradients present in the images, allowing, as we will detail later, a less aggressive subtraction of the sky background, and therefore, higher reliability of the characterisation of the faintest sources. An additional advantage is that the fringing structures contained in the science images are perfectly corrected, allowing to improve the quality of the final images. Given that the sensitivity of the CCD cameras can vary over time, and that the time range between different observations is very wide (of the order of years), the flats are built and applied to data sets taken close in time, typically during the few days of each observation campaign.

Once the images are reduced, we proceed with their combination. First, the images are astrometrically calibrated, using the \software{Astrometry.net} software package \citep{2010AJ....139.1782L} to obtain an approximate solution, and \software{SCAMP} \citep{2006ASPC..351..112B} to obtain the final astrometry. The next step is the coaddition of the individual frames, which is the most crucial step in the process. We perform an iterative loop converging on what we consider the final coadd. The procedure is as follows. First, we obtain a seed coadd, which is the starting point of the iteration. This coadd is produced by subtracting a constant sky value for each of the individual frames to be combined. The consequence of this is that we preserve the lowest surface brightness structures that were not removed by the sky subtraction. However, the frequent gradients in the individual frames produce considerable fluctuations in the sky background which remain in the coadd. This coadd is heavily masked with \software{Noisechisel}, choosing parameters that maximised the masking of the real sources, trying to leave the smooth gradients of the sky background unmasked. This mask produced with the coadd is applied to the individual frames and we perform a smooth polynomial sky fitting to the individual masked frames. We use as sky-fitting surfaces Zernike polynomials \citep[see][]{Zernike34}, always with values equal to or less than n $\leq$ 4, using for each n order all the azimutal components. This Zernike polynomials fitting produce smooth surfaces that do not impact in the oversubstraction around galaxies. Once all the gradients of the individual frames are fitted and subtracted, they are combined to produce a new coadd that is used again as a seed to get a new mask. This process is iterated a number of times. Depending on how strong the gradients are, the number of iterations and the degree of the polynomial are varied to obtain an optimal result. Most cases, given the high quality of flat-fielding, the polynomials have a low degree (n$=2,3$). In extreme cases, and mainly due to the presence of the Moon in the observations, we increase the degree of the polynomial (n$=4$) in order to obtain an optimal final coadd. In general, the reliability of our sky background subtraction allows us to preserve the lower surface brightness structures within our depth limits. We do not find signs of oversubtraction in the images and profiles, such as high-contrast regions close to the end of the galaxies, or systematically truncated profiles (see Sect.\ref{sec:Reliability}). The combination process is performed by photometrically calibrating the individual frames, measuring their signal to noise, and combining them by means of a weighted mean, thus optimising the signal-to-noise ratio of the final coadd.

In Table~\ref{tab:data_sample} we show the final sample. The depth of the images was calculated following the method of \cite{Roman20}, appendix A, according to a standard metric of 3 sigmas in $10^{\prime\prime}\times10^{\prime\prime}$ boxes. We can see that all galaxies have a nominal limiting surface brightness above 29.5\,$\textrm{mag \,arcsec}^{-2}$, except those observed with the JRT, which, as already mentioned, have a slightly lower nominal depth which is compensated by having flatter fields on large angular scales. The total integration time of our campaign, excluding data from the HSC-SSP survey, is 110 hours.

In Fig.~\ref{fig:Comparison} we show the galaxy CIG~340 together with a comparison with SDSS \citep[][]{2009ApJS..182..543A} and Legacy Survey data \citep[][]{2016MNRAS.460.1270D}. The difference in depth is significant between that obtained in our work (30.3\,$\textrm{mag\,arcsec}^{-2}$), SDSS (27.6\,$\textrm{mag\,arcsec}^{-2}$) and the Legacy Survey (28.8\, $\textrm{mag\,arcsec}^{-2}$), all measured at [3$\sigma$; $10^{\prime\prime}\times10^{\prime\prime}$]. While the morphology of CIG~340 appears similar in the SDSS and the Legacy Survey, in our data we detect new structure, with a clear tidal stream to the south of CIG~340 and diffuse light appearing in the direction transverse to the disc, showing a halo-like structure. This highlights the considerable jump in detection power from previously existing data, and the capacity of our observations to reveal the presence of past minor interactions.

\label{sec:DataReduction}

\section{Analysis}
\label{sec:Analysis}

\begin{figure*}[t]
\centering
    \includegraphics[width=\textwidth]{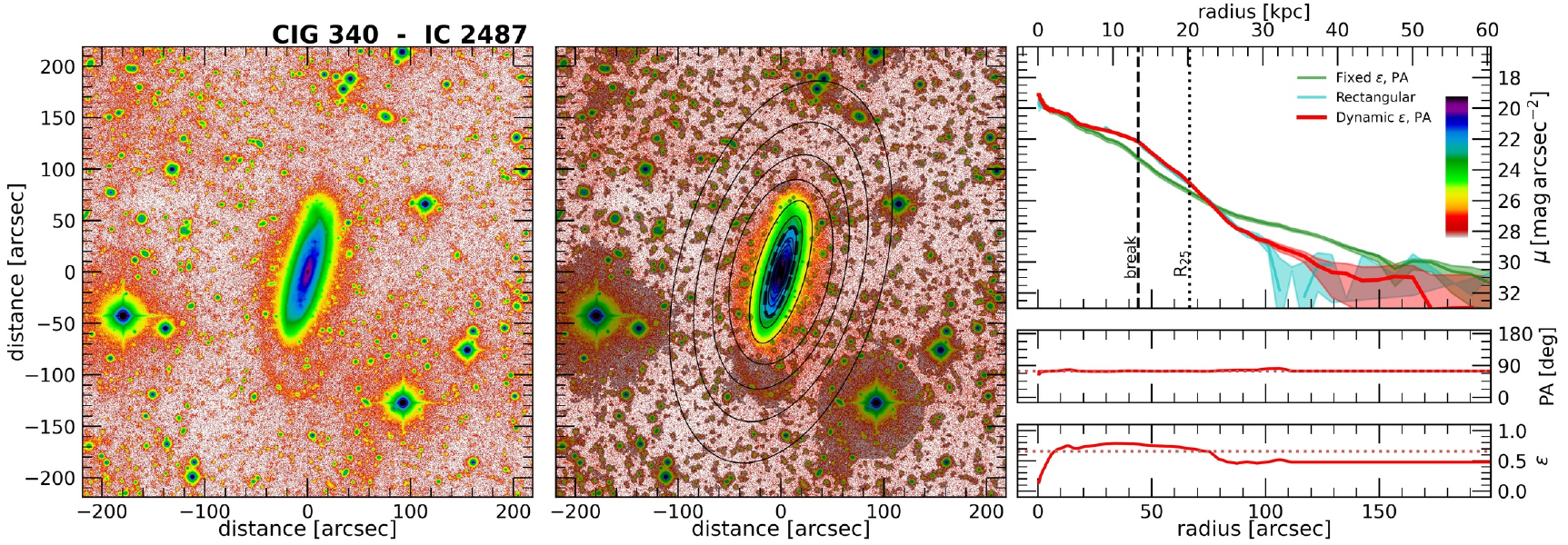}
    \caption{Left panel: r-band image of CIG~340 (IC~2487). The scale of the colours represents the surface brightness of the image and the scale is shown in the profile panel (right) following the y-axis scale. Middle panel: Same image with the mask applied and the elliptical apertures of the dynamic position angle and ellipticity profile. Right panels: Surface brightness profile (top), position angle (middle), and ellipticity (bottom) with respect to the radial distance in arcseconds (bottom) and kpc (top). The red and green curves show the fit with dynamic and fixed elliptical apertures, respectively. The blue curve represents the fit with rectangular apertures. The position of the disc break (if any) is shown with the dashed elliptical aperture and the vertical dashed line in the middle and right panels, respectively. The figures for the whole sample of galaxies are in Appendix~\ref{sec:annexe:Images}.}
    \label{fig:Example_profile}
\end{figure*}

\subsection{Masking procedure}

The masking of light coming from sources other than the target galaxy is one of the most delicate processes needed to obtain the cleanest and deepest possible surface brightness profiles. This task not only requires extracting the signal of astronomical sources from the emission of the background but also requires the attribution (segmentation and deblending) of the signal to one particular source when there is an overlap. 

We combine two of the most popular software tools used in astronomical detection, \software{SExtractor} \citep[][, version 2.25.0]{Bertin1996} and \software{NoiseChisel} and \software{Segment} \citep[][, version 0.18]{segment} part of the \software{GNUastro} package. First, we run \software{NoiseChisel \& Segment} in hot and cold configuration modes for the whole set of images. This allows us to select a region occupied by the galaxy of interest. Since this algorithm can detect very low signal-to-noise ratios (lower than $\lesssim 1$) this region extends to very low surface brightness ($\lesssim 27\,\textrm{mag\,arcsec}^{-2}$), including the outskirts of the galaxy. We execute \software{SExtractor} with the parameters set to optimise the detection of point-like sources (the configuration file can be found in Appendix~\ref{sec:annexe:Config}). All the sources detected outside the galaxy region are then masked.

To improve the detection of the faintest parts of the sources we first smooth the image with the Fully Adaptive Bayesian Algorithm for Data Analysis, \software{FABADA} \cite[][version 0.1]{fabada}. We used an overestimation of the variance of the image in \software{FABADA} to obtain a slightly smoother result and thus larger masks. This allows the detection of even fainter point sources in the proximity of the outskirts of galaxy.

Since we did not mask smaller sources inside the large region occupied by the galaxy we run an extra step to mask these objects. We mask all the sources detected by \software{SExtractor} outside a smaller region, occupied by the galaxy, that corresponds to a level of five times the standard deviation of the image. \software{NoiseChisel} is then run again on the image with the mask applied. This last step allowed the detection of faint extended regions in the image. Given the depth of our data, a considerable amount of faint extended regions (e.g., stellar haloes of background sources, Galactic cirrus, reflections from bright stars, and residual light) appear in the images. All regions that are not spatially connected with the galaxy region and that remain unmasked by the automatic procedure described above are then masked. Finally we visually inspect all masks to improve the masking of sources blended with the target galaxy.


\subsection{Radial profiles} \label{sec:RadialProfiles}

Photometric profiles of galaxies allow regions of approximately equal isophotal magnitude to be averaged to obtain a higher combined signal-to-noise ratio. This allows to reach lower limiting surface brightnesses than with two-dimensional imaging. However, although most galaxies have a simple morphology that allows the different isophotal radii to be fitted by a single elliptical aperture with a given position angle and ellipticity, prominent features in galaxies, such as bars, rings, spiral arms, warps, produce radial variations in the position angle and ellipticity. Additionally, the lower the surface brightness limits of the image, the more galaxies tend to vary their morphology in their outskirts,  as other structures such as outer discs or stellar haloes appear. Thus, the isophotes of galaxies can no longer be modelled by fixed ellipses. We fit elliptical apertures to the image leaving the parameters of the ellipses free in each radial bin \citep[e.g. ][]{Knapen2000,Pohlen06,Mateos15,Pranger17,Watkins19}, thus describing the different structures of the galaxy without any prior assumptions for the whole galactic structure.

We use the implementation in Astropy \citep{astropy13, astropy18} of the iterative ellipse-fitting method described by \cite{Jedrzejewski}. This implementation needs the parameters of a first ellipse to initialise the iterative fitting. We measure the initial parameters using the image moments from a cropped binary image created from the mask image; we select the values above four times the standard deviation of the masked image described in the previous section. This step allows the definition of the morphology in an efficient way. 

We then initialise the elliptical isophote analysis. The elliptical isophote fitting algorithm adjusts ellipses to isophotes of equal intensity pixel values in the images and then computes corrections for the geometrical parameters of the current ellipse by essentially “projecting” the fitted harmonic amplitudes onto the image plane. With this method, we can measure the radial surface brightness profile of the galaxy with a robust mean of the pixels inside the fitted isophotes.

We redefine the morphology of the galaxy using the geometrical parameters of the ellipses as fitted by the algorithm. As a verification step, we produce two other profiles, one using fixed ellipses at the morphology parameters of the galaxy and the other with rectangular apertures separated by a width of five arcseconds along the major axis. These profiles allow us to verify the correct fitting of the previous method by highlighting significant differences in the profiles.

To further improve the reliability of the profiles, we perform additional local sky background corrections. First, we create a preliminary profile with the elliptical apertures parameters fixed. We then select an annular aperture 5 arcsec wide around the galaxy where we reach the level of the local background (often a plateau or an infinite drop). We calculate the sky background value as the mode of the distribution of pixels inside the annular region fitting a Gaussian distribution. This provides a robust sky background reference associated with the galaxy location. In some cases, the surroundings of the galaxy are contaminated by diffuse emission from some other regions, in which case the sky apertures are measured at a larger distance. 

Figure~\ref{fig:Example_profile} shows as example the image of the galaxy CIG~11 (UGC~139). In the left panel we show the surface brightness distribution of the image. In the middle panel we show the same image with a grey layer showing the masked regions. We also show the apertures used for the profiles and the radius where the break is detected (if present). In the right panel we show the surface brightness profile in its three versions: from fixed ellipticity and position angle, with elliptical and rectangular apertures and from elliptical apertures with adaptive ellipticity and position angle. The position of the break (if present) is denoted with a dashed line. In the lower panels we show the variation of ellipticity and position angle as a function of radius. In Appendix \ref{sec:annexe:Images} we show the figures for the rest of the galaxies in our sample.

\subsection{Reliability of the profiles} \label{sec:Reliability}

In order to obtain reliable measurements in the extreme low surface brightness regime, numerous factors have to be taken into account. Most are related to the processing of the data and the instrumentation, such as data reduction and processing, scattered light as described by the point spread function (PSF), or the presence of reflections and artefacts in the images related to instrumentation. Additionally, the presence of Galactic cirrus is ubiquitous and may produce confusion depending on the degree of contamination of the target.

As discussed in Sect.~\ref{sec:data_reduction}, the data processing was carried out using techniques designed to be respectful with the extremely low surface brightness features. This is noticeable in the absence of oversubtraction of the profiles (see Fig. \ref{fig:Example_profile}) in the fainter regions, and allows us to achieve in most cases surface brightness profiles reliable below 30 $\textrm{mag\,arcsec}^{2}$. However, as noted by \cite{2014A&A...567A..97S, 2017A&A...601A..86K, 2019A&A...629A..12M}, and \cite{ 2022ApJ...932...44G}, the light scattered by the bright part of the galaxy itself through the PSF has a decisive impact on the photometric profiles in this extremely low surface brightness regime. In order to obtain better reliability, a proper PSF deconvolution of the galaxy has to be performed. However, given that our data originates from several instruments over a very wide range in time (of the order of years), and that no specific observations of bright stars were carried out we lack a PSF model for each epoch and telescope with which to do the proper PSF deconvolution. Following \cite{Trujillo16} we estimate that the photometric profiles will be unaffected to a surface brightness of around 28\,$\textrm{mag\,arcsec}^{-2}$. Since disc breaks are found to take place at surfaces brightness levels no lower than 26 mag\,arcsec$^{-2}$ \citep[][]{Gutierrez11, Pranger17, Watkins19} our reliability limit is more than sufficient to explore them. However, the lack of adequate PSF models rules out a potential quantitative study of truncations or stellar halos in the outermost regions of the galaxies in our sample.

The presence of Galactic cirrus in our images is also problematic due to the extremely low brightness reached (cirrus indeed appears clearly in some of our images). The maximum possible surface brightness of these cirrus features is 26\,$\textrm{mag\,arcsec}^{2}$ \citep[see][]{Roman20}. Considering that this problem affects mostly surface brightnesses below a value of approximately 26\,$\textrm{mag\,arcsec}^{-2}$ we can again conclude that this does not have an impact on the study of disc breaks. It will, however, have a decisive impact on contaminating the outer regions of galaxies hiding possible interaction features. The presence of cirrus should therefore  be taken into account in the study of the potential presence of minor interactions, as we will describe later on (Sect.~\ref{sec:InteractionsDefinition}).

\subsection{Break identification and classification}

\begin{figure}[t!]
\centering
        \includegraphics[width=0.9\linewidth]{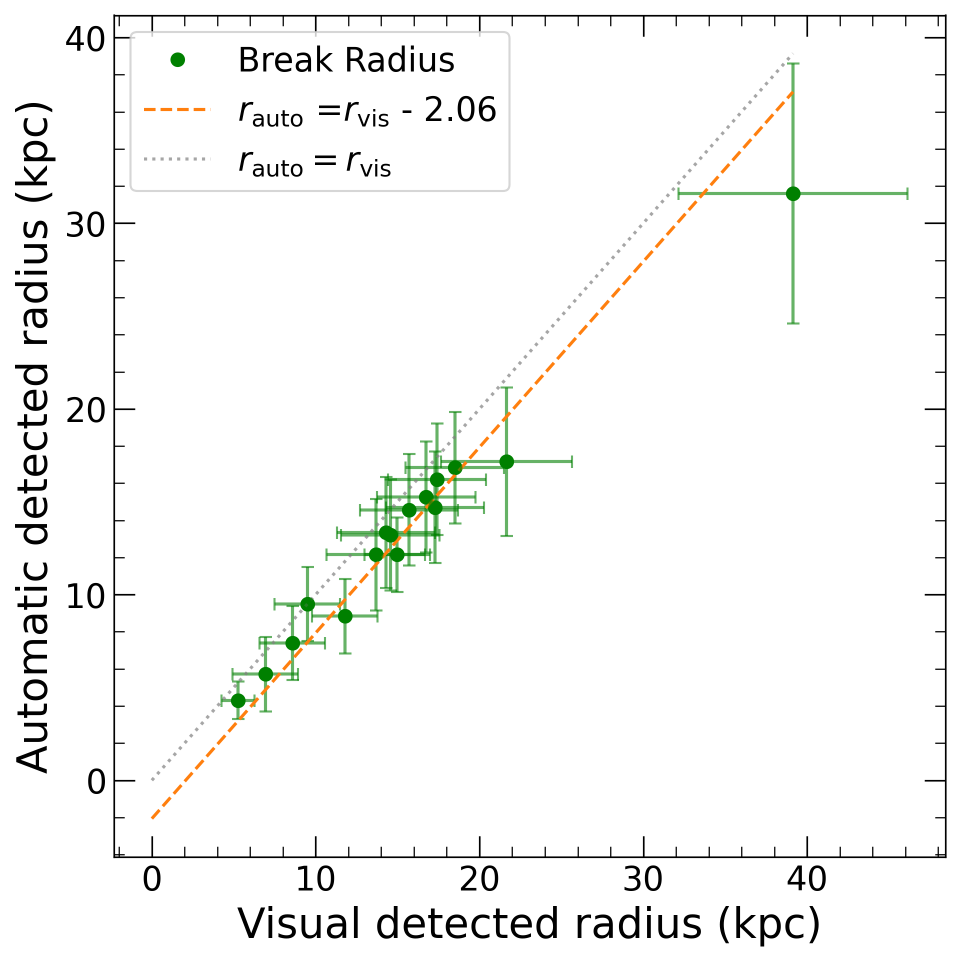}
    \caption{Radii found via an automatic procedure as explained in \cite{Watkins19} compared to those found via visual inspection of the profiles. Gray dotted line represents the 1:1 line, while the orange dashed line represents a linear fit to the data. Error bars were estimated using the distance to the next radial bins in the surface brightness profile at the distance in which the break was measured.}
    \label{fig:Breaks_radii}
\end{figure}

We define a disc break as an abrupt change in the slope of the exponential disc of a galaxy. Following \cite{Erwin05} and \cite{Pohlen06} we define three different types of profiles for our classification: Type~I~$\equiv$~single exponential profile with no change in the slope; Type~II~$\equiv$~down-bending break, with a steeper outer region; and Type~III~$\equiv$~up-bending break, with a steeper inner region. We only search for the most prominent breaks, that are the consequence of a change in the global structure and not due to a local change (such as irregular morphology due to prominent H\,{\sc ii} regions). To characterise the breaks we use two different approaches. First, given the small sample, we classify the break radii and type through a visual inspection of the surface brightness profiles. Second, we use the statistical approach of \cite{Watkins19} where a change point analysis is applied to identify the break radius. This method looks for significant changes in the smooth derivative of the profile. We measure the slope of the surface brightness profile \citep[as done in][]{Pohlen06} using the four nearest points for each radial distance and we smooth the resulting slope with a median filter. We follow a classification similar to that of previous works \citep{Pranger17, Watkins19} for the consistency of the comparison. Fig.~\ref{fig:Breaks_radii} shows the break radii found by the two different methods. We estimate the errors as the resolution of the profiles in the region of the break (distance between each point).  In most cases, both approaches converge to the same solution although the automatic method finds the break $2.06\,\textrm{kpc}$ closer to the galactic centre than visual inspection. We expect an offset, as explained in \cite{Watkins19}, due to the smoothing of the derivative of the profile and the cumulative sum, which can induce small offsets in the radius thus we decide to adopt the value of the radii found by the visual inspection.  In Table~\ref{tab:break_galaxy2} we indicate the classification of the different types of profiles in our sample, together with the values of the break radii, found visually, and the surface brightness levels where the breaks occur. 

\begin{table}[t!]
\caption{Results of the break identification.}
\centering
\resizebox{\columnwidth}{!}{%
\begin{tabular}{lllcccc}
\hline \hline\\[-8pt]
Galaxy & Break &  Break & Break & $Q_{\textrm{Kar,p}}$ & $\eta_{k,\textrm{p}}$  & Class \\
& Type & Radius & $\mu$ & & & \\
\hline \hline\\[-8pt]
CIG 11   & II& $19\pm3$ & $24.8\pm0.7$  & -1.23 & 2.03 & - \\
CIG 33   & III& $12\pm2$ & $23.6\pm0.4$  & -3.11 & 2.09 & H \\
CIG 59   & I & - & - &  -3.10 & 1.89 & - \\
CIG 94   & II& $10\pm2$ & $22.5\pm0.4$ & - & - & - \\
CIG 96   & III& $17\pm3$ & $25.5\pm0.2$ & - & - & H \\
CIG 100  & II & $5\pm1$ & $21.9\pm0.2$ & -3.01 & 2.16 & - \\
CIG 154  & I & - & - &  - & - & C \\
CIG 279  & I & - & - &  -3.04 & 2.16 & - \\
CIG 329  & II& $16\pm3$ & $22.0\pm0.3$ & -3.57 & 1.31 & T \\
CIG 335  & II& $15\pm3$ & $23.5\pm0.5$ & -3.48 & 1.42 & H \\
CIG 340  & II& $14\pm3$ & $22.4\pm0.4$ & -3.85 & 1.04 & H+T \\
CIG 512  & II& $9\pm2$ & $24.2\pm0.5$ & -3.40 & 1.38 & - \\
CIG 568  & II& $22\pm5$ & $23.3\pm0.5$ & -2.80 & 2.44 & - \\
CIG 613  & II& $39\pm7$ & $24.9\pm0.9$ & -3.33 & 1.78 & T \\
CIG 616  & I & - & - &  -3.02 & 2.55 & H \\
CIG 626  & I & - & - &  -3.93 & 1.05 & - \\
CIG 744  & I & - & - &  -2.20 & 2.47 & - \\
CIG 772  & II& $17\pm3$ & $23.1\pm0.4$ & - & - & - \\
CIG 800  & II& $17\pm3$ & $23.6\pm0.5$ & - & - & C \\
CIG 838  & I & - & - &  - & - & - \\
CIG 947  & I & - & - &  - & - & C \\
CIG 971  & I & - & - &  - & - & C \\
CIG 1002 & II& $14\pm3$ & $22.8\pm0.5$ & -2.42 & 2.69 & - \\
CIG 1004 & II& $15\pm3$ & $22.3\pm0.4$  & -2.25 & 1.49 & C \\
CIG 1047 & II& $7\pm2$ & $22.6\pm0.9$ & -2.62 & 2.25 & H \\
\hline 
\end{tabular}
}
\tablefoot{The first column shows the galaxy CIG name. The second, third and fourth show the type of break found, the break radii in kpc if any, and the surface brightness level in mag\,arcsec$^{-2}$ of the break, respectively. The fifth and sixth columns show the interaction parameters $Q_{\textrm{Kar,p}}$ and $\eta_{k,\textrm{p}}$ from the AMIGA catalogue (\cite{Argudo-Fernandez13}). The seventh column indicates the presence of interaction features found in this work. We define three different categories, Halo-perturbed (H), Tidal streams (T) and Cirrus contamination (C). More information about the classification can be found in Sect.~\ref{sec:InteractionsDefinition}.}
\label{tab:break_galaxy2}
\end{table} 

\subsection{Identification of interactions}
\label{sec:InteractionsDefinition}

We visually inspect our sample in search of signatures of perturbations following the definitions of \cite{2023A&A...671A.141M}, and references therein. The large field of view of our images corresponds to at least 100\,kpc, enough to explore the presence of interactions with confidence. In order to provide a simple and general classification with which to introduce a perturbation parameter, we distinguish galaxies according to their morphology as follows. Tidal Stream (T): The galaxy shows a tidal stream, elongated in shape consistent with an in-falling satellite in current interaction. Halo-Perturbed (H): The galaxy has asymmetries or debris in the outermost part of the disc. Cirrus (C): In the case of strong Galactic cirrus contamination, a correct interpretation of the degree of galaxy perturbation or interaction is not possible. In this case, we discard the galaxy as non-tractable, excluding it from the statistics that imply using the presence of interactions. When a galaxy appears fairly symmetrical in morphology with no signs of disturbance, no classification is given. 
We describe the presence of interactions as effects from mergers in the outer regions of galaxies. These are distinguishable from lopsidedness potentially produced by gas \citep{2017ASSL..434..209B}, since our surface brightness limits allow us to explore the outer or halo regions of galaxies, beyond simply tracing the central morphology where star formation is dominant.

\begin{figure*}[]
\centering
        \includegraphics[width=1.0\textwidth]{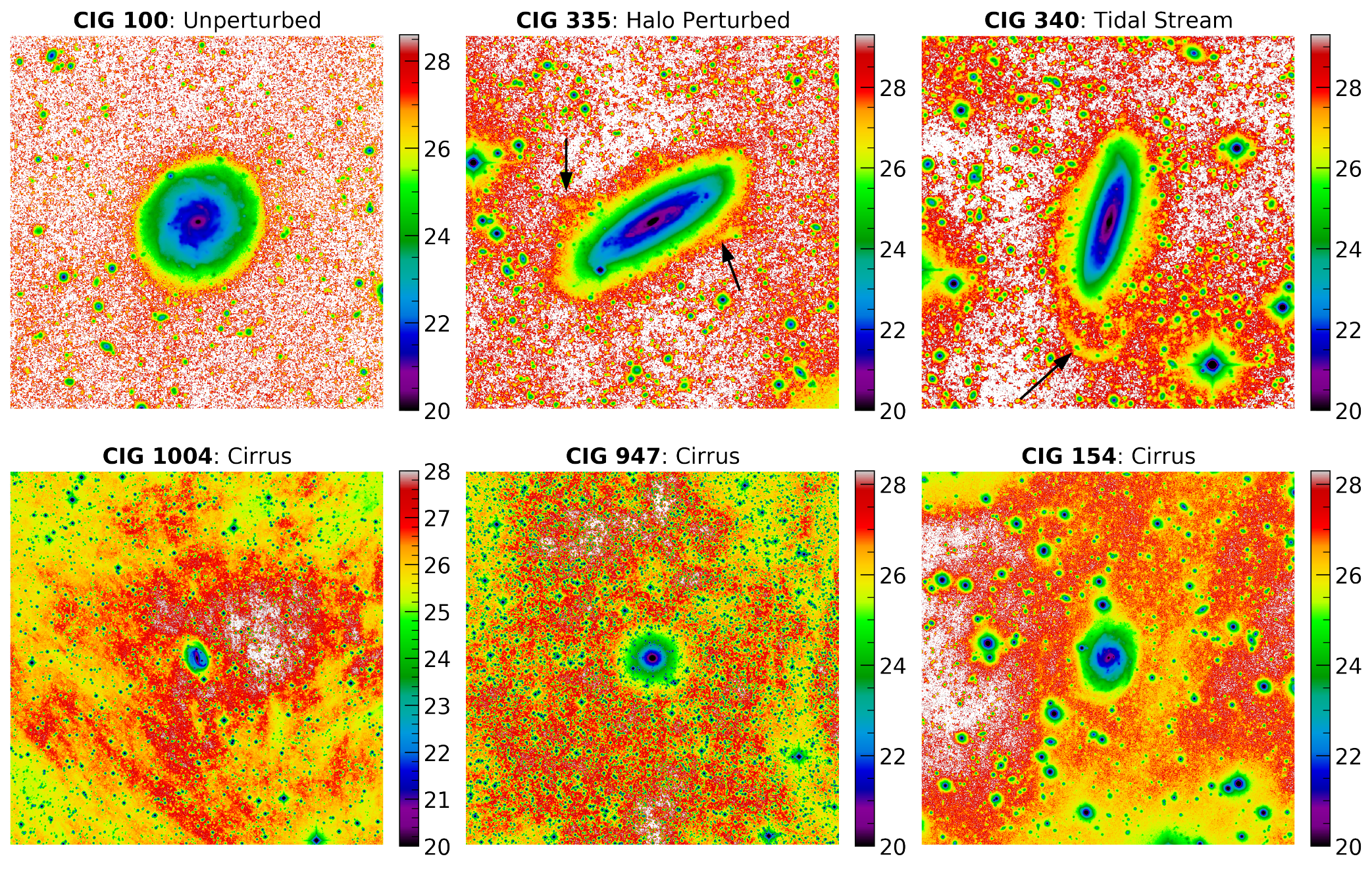}
    \caption{Different examples of the interaction classification scheme. In the top panel, we show an unperturbed, symmetric example (top-left panel), CIG~100 (UGC~1863). The top middle and right panels show examples of perturbed galaxies. CIG~335 (NGC~2870) shows an overdensity in the halo region, while CIG~340 (IC~2487) shows a tidal stream in the southern region. All perturbations features are indicated with a black arrow. The bottom panels represent galaxies classified as cirrus-contaminated. CIG~1004 (NGC~7479), CIG~947 (NGC~7217), and CIG~154 (UGC~1706) are shown in the bottom left, middle and right panels, respectively. These examples show considerable amounts of diffuse emission associated with Galactic cirrus only seen in our images. All of the images, except the one of CIG~100, were smoothed with \software{FABADA} to enhance the structure.}
    \label{fig:Panel_interacciones}
\end{figure*}

The result of this classification for each galaxy is shown in Table~\ref{tab:break_galaxy2}, and in Fig.~\ref{fig:Panel_interacciones} we show representative examples of the interaction classification scheme. In the top-row panels, we show a symmetric non-classified (left), halo-perturbed (middle), and tidal stream (right) examples. In the bottom panels we show three different examples of galaxies with strong Galactic cirrus contamination that do not allow us to assure the presence or absence of potential interaction features in the galaxies.

 \section{Results} \label{sec:Results}
 
We present below individual comments on each of the galaxies with a short discussion of the decisions made in the different classifications. In Table~\ref{tab:break_galaxy2} we show the results of the classification of profiles for each galaxy, together with the radius at which a break has been found (if any) and its surface brightness level.

\begin{itemize}
    \item[*] \textbf{CIG~11 [UGC~139]}: Although this galaxy could have been classified as Type~III+II at $10 + 20\,\textrm{kpc}$ \citep[see discussion by][]{Watkins19}, we classify as a Type~II. The reason is that extra flux originates by H\,{\sc ii} regions seen in the inner disc region. This extra flux could be misinterpreted as Type~III, although according to \cite{Pranger17} Type~III breaks occur further away. Thus, for the sake of a fair comparison, we classify this as a Type~II break at $19\,\textrm{kpc}$. We also consider this galaxy as unperturbed due to its symmetric structure. Diffuse cirrus emission is present, but that does not prevent the detection of interactions.
    
    \item[*] \textbf{CIG~33 [NGC~237]}: We consider this a clear Type~III with a break radius of  $\sim$12~kpc and a perturbed halo. The diffuse emission around the galaxy is asymmetric with a bump in the North-East region of the outskirts. 
    
    \item[*] \textbf{CIG~59 [UGC~1167]}: Exponential disc without break, with symmetric structure. Presence of instrumental reflection of a nearby star that when correctly masked does not contaminate the rest of the galaxy.
    
    \item[*] \textbf{CIG~94 [UGC~1706]}: Type~II break at $\sim10\,\textrm{kpc}$, with absence of interaction signatures. 
    
    \item[*] \textbf{CIG~96 [NGC~864]}: Type~III break at $\sim16\,\textrm{kpc}$ with perturbed halo. Extra flux at the North-East region of the disc outskirts.
    
    \item[*] \textbf{CIG~100 [UGC~1863]}: Type~II break at 5\,kpc, symmetric structure. Possible Type~III break around ~18.5\,kpc at 28 mag\,arcsec$^-2$. However, at this depth, we cannot distinguish between instrumental effects such as extended PSF contribution as explained in Sect. \ref{sec:Reliability}.
    
    \item[*] \textbf{CIG~154 [UGC~3171]}: Exponential disc with symmetric structure. Hints of a larger spiral arm, at around $\sim$ 17\,kpc,  of the southern region. The presence of filamentary cirrus prevents us from determining the presence of any faint interaction feature in the outer regions of the galaxy.
    
    \item[*] \textbf{CIG~279 [NGC~2644]}: Exponential disc with symmetric structure. The presence of diffuse cirrus emission would not prevent the detection of interaction features if present. Possible Type~III at $\sim13\,\textrm{kpc}$, however, the depth and presence of cirrus do not allow us to distinguish between the possible origins. 
    
    \item[*] \textbf{CIG~329 [NGC~2862]}: Type~II break at $\sim16\,\textrm{kpc}$, clear tidal stream at the end of the disc in the South-East region. Presence of instrumental reflection of a nearby star that correctly masked does not contaminate the rest of the galaxy.
    
     \item[*] \textbf{CIG~335 [NGC~2870]}: Type~II break at $\sim15\,\textrm{kpc}$. Signatures of overdensities and perturbations in the halo region so we classify this as a perturbed halo. Fig.~\ref{fig:Panel_interacciones} shows an enhanced image of the galaxy for greater clarity. 
     
    \item[*] \textbf{CIG~340 [IC~2487]}: Type~II at $\sim14\,\textrm{kpc}$, clear tidal stream in the south region. 
    
    \item[*] \textbf{CIG~512 [UGC~6903]}: Symmetric galaxy with a Type~II break at $\sim9\,\textrm{kpc}$. Oversubstractions effects could cause the significant drop in brightness seen at 20\,kpc, which can be a result of the Subaru pipeline.  
    
    \item[*] \textbf{CIG~568 [UGC~8170]}: Symmetric galaxy with a Type~II break at $\sim21\,\textrm{kpc}$.
    
    \item[*] \textbf{CIG~613 [UGC~9048]}: Type~II break at $\sim40\,\textrm{kpc}$. Galaxy also exhibits signatures of a very faint and warped tidal stream. Appendix~\ref{sec:annexe:Images} shows an enhanced image with an arrow showing this stream.
    
    \item[*] \textbf{CIG~616 [UGC~9088]}: Type~I exponential disc, with an elliptical-like shape in the halo region with extra flux perpendicular to the major axis. Extra flux in clumps in the outermost regions indicating past interaction. Classified as perturbed halo. 
    
    \item[*] \textbf{CIG~626 [NGC~5584]}: Symmetric galaxy with a featureless disc profile, Type~I. 
    
    \item[*] \textbf{CIG~744 [UGC~10437]}: Although this galaxy seems to have some fluctuations between $7\,\textrm{kpc}$ and $15\,\textrm{kpc}$ that could be classified as breaks, the constant global decrease in average surface brightness in the disc makes us to classify it as an exponential disc (Type~I). We identify the local fluctuations coming from clumpy H\,{\sc ii} regions of the arms. The absence of features indicating interactions makes us classify this galaxy as symmetric.
    
    \item[*] \textbf{CIG~772 [IC~1231]}: Symmetric galaxy with Type~II break at $17\,\textrm{kpc}$.
    
    \item[*] \textbf{CIG~800 [NGC~6347]}: Type~II break at $\sim19\,\textrm{kpc}$. The galaxy is fully embedded in a region heavily contaminated by cirrus, making the classification of possible interaction features unfeasible. Furthermore, several stars lay in the line of sight, in the North-West region, and faint contamination due to the extended PSF can cause extra light at the end of the profile (see Sect. \ref{sec:Reliability}).
    
    \item[*] \textbf{CIG~838 [IC~1269]}: Galaxy with an exponential Type~I disc. From our optical imaging or the derived profile we do not detect any signs of interaction, even though the northern spiral arm shows asymmetries with respect to the other one. Some diffuse cirrus emission is present, but not enough to prevent the detection of faint interaction features. 
    
    \item[*] \textbf{CIG~947 [NGC~7217]}: Although this object can be misinterpreted as an early-type galaxy, high-resolution colour images, show spiral structures with two blue rings in the inner and outer regions. The radial profile exhibits a clear exponential disc. We see clear signatures of cirrus contamination (Fig.~\ref{fig:Panel_interacciones}) preventing any detection of possible interaction features.
    
    \item[*] \textbf{CIG~971 [UGC~12082]}: Exponential disc, Type~I. The galaxy is fully embedded in a region heavily contaminated by cirrus, making the classification of possible interaction features unfeasible. Furthermore, the field is crowded with stars.  This contamination can be also be seen in the rectangular radial profile as a plateau at $\sim28\,\textrm{mag\,arcsec}^{-2}$ (as explained in Sect. \ref{sec:Reliability}).
    
    \item[*] \textbf{CIG~1002 [NGC~7451]}: Type~II break at $14\,\textrm{kpc}$. We detect a possible companion in the South-West region of the galaxy. However, there is no available spectra to confirm the association of the two galaxies. We do not see any interaction signature between them. 
    
    \item[*] \textbf{CIG~1004 [NGC~7479]}: Type~II with a break radius of $15\,\textrm{kpc}$.
    The difference between the fixed and free ellipticity surface brightness profiles comes from the presence of the extended stellar bar. We are also able to see the same break with the rectangular aperture. The galaxy is fully embedded in a region heavily contaminated by cirrus (see Fig.\ref{fig:Panel_interacciones}), making the classification of possible interaction features unfeasible from deep optical imaging. The asymmetry of the galaxy disc and spiral arms is well known, however, and may indicate a minor merger origin \citep[e.g.][]{1998MNRAS.297.1041L}.
    
    \item[*] \textbf{CIG~1047 [UGC~12857]}: Type~II break at $8\,\textrm{kpc}$. We see a warp in the southern part of the outer disc, so we classify this galaxy as a perturbed halo. Presence of diffuse cirrus emission that would not prevent the detection of interaction features if present.

\end{itemize}
 \begin{figure*}[ht!]
\centering
   \includegraphics[width=\textwidth]{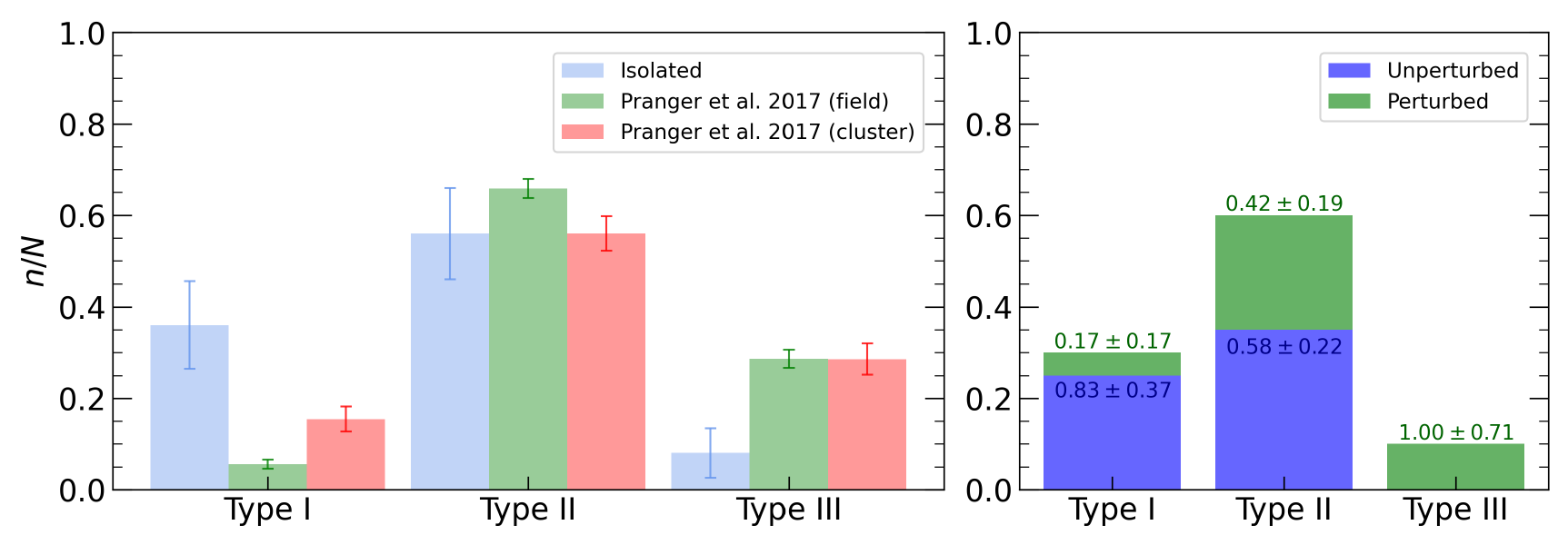}
    \caption{Left: Normalised distribution of the frequency of the types of breaks found in this work (left-blue) and in \cite{Pranger17} for their field (middle-green) and cluster (right-red) samples. Right: Normalised distribution of the frequency of the types of breaks. Galaxies strongly contaminated by cirrus are excluded from this figure. We also show the contribution of unperturbed (blue) and perturbed (green) for galaxies with each type of break.}
    \label{fig:histogram}
\end{figure*}
In total, among 25 galaxies, we identify 9~Type~I single exponential with no significant break, 14~Type~II down-bending breaks, and 2~Type~III up-bending breaks. The overall statistics of the sample are $36\% \pm 10\%$~Type~I, $56\% \pm 10\%$~Type~II, and $8\% \pm 5\% $~Type~III breaks. We estimate the uncertainties assuming a binomial distribution, $\epsilon = \sqrt{f\,(1-f/N)}*(100/N)$ [\%] where $f$ is the fraction of galaxies within each type and $N$ is the total number of galaxies in our sample ($N=25$).

\subsection{Break type vs. interactions}

We find 5 galaxies with strong contamination by Galactic cirrus clouds that do not allow us to confirm or rule out the presence of interactions thus these 5 galaxies are excluded from this analysis. Among the remaining 20, we find 8 galaxies with signatures of interactions. Of these,  4 show asymmetries in the halo, 3 have tidal streams, and the remaining one shows both tidal streams and halo asymmetries. Therefore, 40\%$\pm$14\% of the isolated galaxies in our classified sample show the presence of interactions; out of these galaxies, 25\%$\pm$10\% show an asymmetric halo and 20\%$\pm$8\% show some tidal streams. We found 12 galaxies (60\%$\pm$17\%) with no interaction features. 

Given the small sample size, we consider any type of interaction as a single class named perturbed (8 galaxies) (presence of interactions) while the remaining galaxies (that lack cirrus contamination) are classified as unperturbed (12 galaxies). In the right panel of Fig.~\ref{fig:histogram} we show the fraction of perturbed and unperturbed galaxies for each type of break. We found that for Type~I breaks, 6 ($83\% \pm 37\%$) are unperturbed galaxies and 1 ($17\% \pm 17\%$) is perturbed; for Type~II, 8 ($58\% \pm 22\%$) are unperturbed and 5 ($42\% \pm 19\%$) are perturbed. Finally, we find that the two galaxies with Type~III are perturbed ($100\% \pm 71\%$). We estimate the errors assuming Poisson statistics since we have few of galaxies in each type. 

The low number of galaxies in our sample is insufficient to allow us to make strong statements. However, we find certain indications that are undoubtedly interesting. First, both Type~III galaxies (CIG~33 and CIG~96) appear strongly perturbed. This is noticeable in the images in the Appendix in which we can appreciate how these two galaxies have an expanded halo with clear signs of strong disturbance, possibly due to a recent major merger. Second, we find a significantly higher fraction of perturbed galaxies (42\% $\pm$ 19\%) among Type~II break hosts than among Type~I (17\% $\pm$ 17\%). 

\subsection{Break type vs. environment}

In the left panel of Fig.~\ref{fig:histogram} we show the fraction of surface brightness profile types in our sample of isolated galaxies in comparison with the previous work by \cite{Pranger17} who explored the surface brightness type in a sample of 700 disc galaxies at low redshift ($z < 0.063$) using SDSS data. These galaxies were classified according to their environment into field (low-density) and cluster (high-density). The fractions of Type~I, II, and III discs found by \cite{Pranger17} are 29 ($6\%\pm1\%$), 343 ($66\%\pm2\%$), and 149 ($29\%\pm2\%$) in the field sample and 27 ($15\%\pm3\%$), 98 ($56\%\pm4\%$), and 50 ($29\%\pm3\%$) in the cluster sample. For the sake of a fair comparison, we estimate the \citet{Pranger17} uncertainties in the same way as we do for our results. We compare these statistics with the sample of isolated galaxies from our work. As illustrated in Fig.~\ref{fig:histogram}, left panel, we find a considerably higher fraction of isolated Type~I galaxies and a lower fraction of isolated Type~III galaxies than what was found by \cite{Pranger17} for field galaxies and clusters. The results for Type~II are similar for different environments. 

A Kolmogorov–Smirnov statistical test proves that our results are significantly different from those of \cite{Pranger17} ($P$-value < 0.01). The highest difference between our results and those of the \cite{Pranger17} sample is found for Type~I discs. We find six times  more single exponential discs in our isolated sample. This difference holds when compared to other studies, such as that by  \cite{Gutierrez11}, who found around $\sim 10-15 \%$ of Type~I disc in field late-type spirals, around two to three times less often than us. We also find a much lower fraction, around seven times lower, of Type~III discs than \cite{Pranger17}.

\begin{figure}[]
\centering
        \includegraphics[width=\columnwidth]{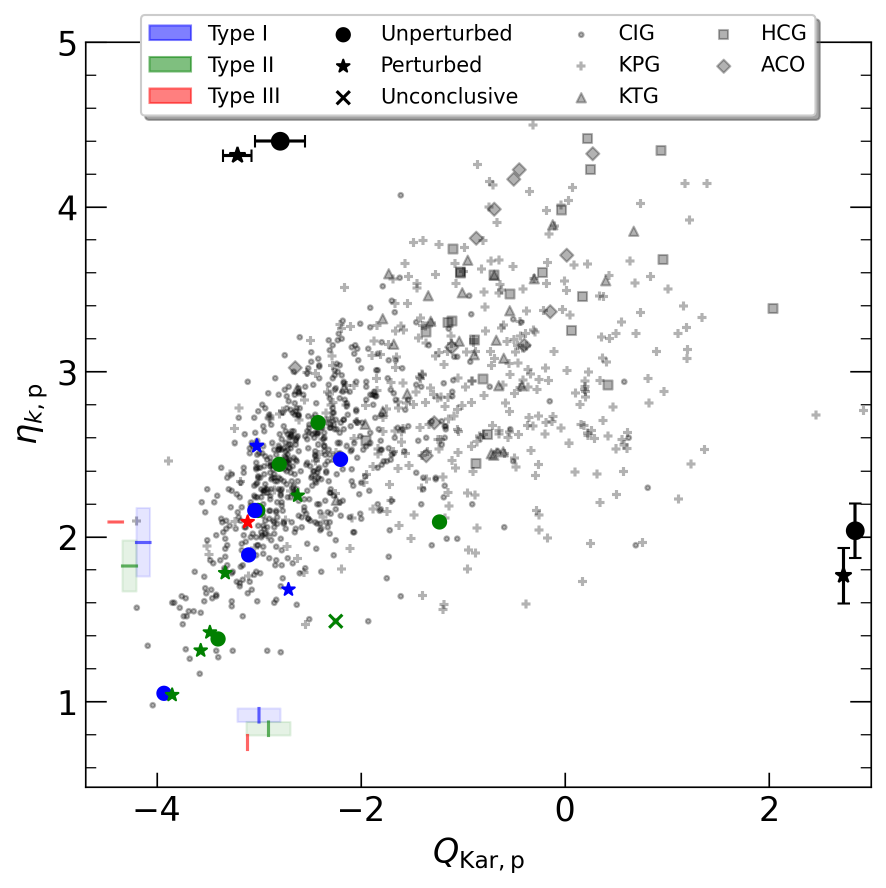}
    \caption{Photometric isolation parameter for 18 galaxies of our sample, the local number density of neighbour galaxies, $\eta_{k,\textrm{p}}$ compared to the tidal strength, $Q_{\textrm{Kar,p}}$ from \cite{Argudo-Fernandez13}. The colour code represents the type of disc found, blue, green, and red for Types~I, II, and III, respectively. The mean values of each type are represented by the line close to the axis (bottom and left) with its corresponding error. Circle and star symbols represent unperturbed and perturbed galaxies, respectively. The mean values of perturbed and unperturbed galaxies are represented by the symbols close to the axis (upper and right) with its corresponding error. The cross symbols correspond to galaxies not classified due to galactic cirrus. Background grey values represent Galaxies from different samples (see Sect.~\ref{sec:Results} for details). }
    \label{fig:Dens_vs_break}
\end{figure}

In Fig.~\ref{fig:Dens_vs_break} we show a correlation test between the degree of isolation and the disc type of the galaxies in our sample. We plot the local number density of neighbours galaxies $\eta_{k,\textrm{p}}$ and the tidal strength $Q_{\textrm{Kar,p}}$ obtained by \cite{Argudo-Fernandez13} for the 18 galaxies in our sample that have these parameters calculated. An increasing value for $\eta_{k,\textrm{p}}$ or $Q_{\textrm{Kar,p}}$ indicates a higher environmental density.  We show as a comparison these parameters for other galaxy catalogues, including galaxies located in regions of higher density. In particular, we show values for isolated pairs of galaxies \citep[KGP;][]{Karachentsev72}; galaxy triplets \citep[KTG;][]{Karachentseva79}; galaxies in compact groups \citep[HCG;][]{Hickson82}; and galaxies in Abell clusters \citep[ACO;][]{Abell58,Abell89} computed by \cite{Argudo-Fernandez13}. We find that the galaxies in our sample tend to be located at low values of $\eta_{k,\textrm{p}}$ and $Q_{\textrm{Kar,p}}$, confirming their location in regions of low environmental density, a consequence of the sample selection criteria. The colour code indicates the type of surface brightness profile, while the symbols show the two populations according to our classification, perturbed and unperturbed galaxies. The average values of the $\eta_{k,\textrm{p}}$ and $Q_{\textrm{Kar,p}}$ parameters for each break type are indicated by colour bars, while for the unperturbed and perturbed population, they are shown with symbols. We see that for the galaxies in our sample, the different disc types tend to be located on average in regions of similar density, with these regions in any case having a very low density.

\section{Discussion}
\label{sec:Discussion}

We present deep optical imaging of a sample of 25 isolated galaxies from the AMIGA project in order to reach low surface brightness limits. The nominal surface brightness limits achieved are $\mu_{r,\textrm{lim}}\,>\,29.5\,\textrm{mag\,arcsec}^{-2}$~[3$\sigma$;\,$10^{\prime\prime}\times10^{\prime\prime}$], and $\mu_{L,\textrm{lim}}\,>\,28.5\,\textrm{mag\,arcsec}^{-2}$~[3$\sigma$;~$10^{\prime\prime}\times10^{\prime\prime}$] for the three galaxies observed with the JRT. Because of our careful data processing, our images show an absence of oversubtracted regions and a high efficiency in the detection of extreme low surface brightness features.

The depth of our data is more than 1 $\textrm{mag\,arcsec}^{-2}$ (\textit{r} band) deeper than that in preceding studies \citep[see e.g., ][using the Legacy Survey]{2023A&A...671A.141M}. This gives us the possibility to look for the presence of minor interaction signatures at very low surface brightness. A representative example of this quantitative leap is shown in  Fig.~\ref{fig:Comparison} where we compare images of CIG~340 (IC~2487) with SDSS and Legacy Survey data. This comparison clearly shows that only in our data is it possible to detect the clear interaction that CIG~340 is undergoing with a low-mass satellite, appearing as a tidal stream with surface brightness around $26.8-27.3 ~ \textrm{mag\,arcsec}^{-2}$ and an extension of around $\sim 116\,\textrm{arcsec}$ ($35\,\textrm{kpc}$), not detectable in shallower optical data.

The potential detection of minor interactions has a decisive impact on the interpretation of galaxy properties, and the isolated galaxy CIG~340 is a perfect example. Shallow optical images of CIG~340 revealed a fairly symmetric disc, albeit with a small disc warp. Additionally, the H\,{\sc i}-integrated spectra from single-dish observations showed a very symmetric profile \citep[see][]{2011A&A...532A.117E}. More recent high-resolution interferometric observations by \cite{Scott14} revealed a striking asymmetry of the H\,{\sc i} component of CIG~340, with 6\% of the H\,{\sc i} mass located in an extension of the disc to the north. These findings led \cite{Scott14} to propose two different hypotheses to explain them. On the one hand, this H\,{\sc i} asymmetry could be caused by a minor interaction with a satellite, not detected in the optical images available at the time. On the other hand, it could be due to some internal secular process, for example the result of a long-lived dark matter halo asymmetry. More recent work by \cite{Kipper20} proposed that the gravitational interaction of a background medium of dark matter particles in the surroundings of CIG~340 is capable of inducing a dynamical friction enough to cause the H\,{\sc i} asymmetries observed by \cite{Scott14}. In light of the results of our work, we can affirm that the H\,{\sc i} asymmetries observed in the isolated galaxy CIG~340 are most likely caused by a minor interaction, the signatures of which we have unveiled for the first time. \\
The galaxy CIG~96 (NGC~864) is another well-studied isolated galaxy. Recent work detected two H\,{\sc i} asymmetries in the North-West and South-East regions of the galaxy \citep{Ramirez-Moreta18}, not detected in the optical range. The main hypotheses to explain the HI features of CIG~96 are possible accreted companions and cold gas accretion. However, \citet{Ramirez-Moreta18} ruled out the possibility of major mergers events due to the isolation criteria (discarded possible interactions in the last 2.7\,Gyr). Despite the high degree of isolation of this galaxy, with our deep data we are able to detect a bigger external faint halo in the galaxy, which along with its Type~III profile suggests that this galaxy might have experimented a recent merger event causing the extended emission of light \citep[e.g.][]{2001MNRAS.324..685L,Lotz08,2014A&A...570A.103B,2015A&A...582A..21B,2022MNRAS.509..261P}. \\
The possibility of achieving surface brightness levels as low as the ones shown here offers a new decisive parameter to explain some morphological features in galaxies. This would not only be useful to investigate the main reasons for H\,{\sc i} asymmetries of galaxies, but also in other fields such as the possible induction of active galactic nuclei (AGN) by minor interactions in galaxies, among many others. These issues will be investigated in future works.

The types of discs in the isolated galaxies from our sample show significant differences with respect to the results of previously studied samples in other environments. We find a significantly higher fraction of Type~I and a significantly lower fraction of Type~III profiles than in works by \cite{Pranger17} and \cite{Gutierrez11} for denser environments. This is a striking result which we further discuss now.

Type~III profiles can be produced by mergers of galaxies \citep[e.g., ][and references therein]{2001MNRAS.324..685L,2014A&A...570A.103B, 2022MNRAS.509..261P}. The low number of Type~III profiles in our sample of isolated galaxies is in agreement with this statement since undoubtedly a lower density would imply a lower merger ratio. Additionally and importantly, the only two Type~III galaxies found in our sample (CIG~33 and CIG~96) show disturbed morphology with a puffed-up external halo, compatible with a recent major merger \citep{Lotz08, 2015A&A...582A..21B}. 

Type~II galaxies are widely agreed to be the result of discs with breaks caused by a star formation threshold \citep[e.g., ][]{, 2020ApJ...897...79T, 2022MNRAS.509..261P}. A cessation or decrease in star formation would tend to homogenise through stellar migration effects \citep[][]{SB09}  the stellar populations and produce a Type~I profile \citep[e.g., ][]{2022MNRAS.509..261P}. In our case this is hardly testable directly due to the absence of star formation measurements in our sample. However, an indirect hint that could indicate that this is the case is the considerably higher fraction of perturbed Type~II (42\%$\pm$19\%) galaxies when compared to Type~I (17\%$\pm$17\%) galaxies. While the rate of star formation may depend on various circumstances such as the rate of pristine gas inflow, it is known that satellite interactions are capable of triggering star formation \citep[][]{2000ApJ...530..660B, 2004MNRAS.352.1081A, 2015MNRAS.454.1742K, 2021MNRAS.501.1046M}, which would be in accordance with our findings. 
The  low ratio of perturbations detected in Type~I discs galaxies in our results and the lower specific star formation rate in isolated galaxies than in higher-density environments \citep[][]{2011A&A...534A.102L,2015MNRAS.451.1482M, Cavity} suggest that unperturbed galaxies, evolving slowly with a low star formation rate could explain the high rate, of Type~I discs in isolated galaxies.

\section{Conclusions}
\label{sec:Conclusions}

We study a sample of 25 ``isolated'' galaxies from the AMIGA revised CIG catalogue, which lack major companions, to identify how internal or external processes impact the discs of galaxies. We conduct a diverse observational campaign using the INT, VST, JRT, and archival data from HSC-SSP to obtain unprecedentedly deep images. We measure the surface brightness profiles and classify the galaxies according to their disc break (Type~I $\equiv$ single exponential, Type~II $\equiv$ down-bending, Type~III $\equiv$ up-bending) and to the presence of interaction signatures (Tidal Stream, Halo perturbation, and unperturbed). The conclusions of our work are the following:
\begin{itemize}
    \item Our images have similar depth as those to come from future surveys, like LSST, through careful data processing and background subtraction. The nominal surface brightness limit of the images is $\mu_{r,\textrm{lim}}\,>\,29.5\,\textrm{mag\,arcsec}^{-2}$~[3$\sigma$;\,$10^{\prime\prime}\times10^{\prime\prime}$]. The data processing is optimized to preserve low surface brightness features. 
    \item As a result of the depth obtained, we can trace interaction signatures in galaxies classified as isolated (see Figure~\ref{fig:Comparison}). However, five galaxies are affected by cirrus and we could not explore the presence of signatures, ruling them out from this analysis. We find that 25\% $\pm$ 10\% of the galaxies in our sample show an asymmetric halo and 20\% $\pm$ 10\% show a tidal stream (see Figure~\ref{fig:Panel_interacciones}). In total, 40\% $\pm$ 14\% show signs of interactions. 
    \item We are able to produce reliable surface brightness profiles down to a critical surface brightness of $\gtrsim 30\, \textrm{mag\,arcsec}^{-2}$ for all of the galaxies in our sample (see Appendix~\ref{sec:annexe:Images}).
    \item We successfully classified the disc type in all the galaxies in our sample. We identify nine ($36\% \pm 5\%$) Type~I discs with no significant break, fourteen ($56\% \pm 7\%$) Type~II down-bending discs, and two ($8\% \pm 3\% $) Type~III up-bending discs. 
    \item The fraction of perturbed galaxies correlates with the type of disc. We identify $17\%\pm17\%$, $42\%\pm19\%$ , and $100\%\pm71\%$ perturbed galaxies with Type~I, II, and III discs, respectively. 
    \item We find significantly higher Type~I and lower Type~III frequencies with respect to other studies \citep[such as][]{Pranger17,Gutierrez11} with more perturbed galaxies in Type~II and III than in Type~I. This is in agreement with a proposed formation scenario in which Type~III discs are formed via interactions such as major mergers \cite[see][]{Laine14, Watkins19}, and Type~II discs stem from a star formation threshold. The increased fraction of Type~I discs with respect to that in other samples could be attributed to the low ratio of disturbance in our sample of isolated galaxies. 
\end{itemize}

In the near future, the advent of the next generation optical and infrared surveys (e.g., LSST, Euclid) will increase the number of galaxies observed at surface brightness limits equivalent to those that we present in this work by several orders of magnitude. This will allow to further strengthen and refine our statements, provided the data reduction and analysis allows for the detection of low surface brightness.

\begin{acknowledgements}
We thank Ignacio Trujillo for helpful insights about this work and Aaron Watkins for providing us with the implementation of the automatic break detection method. 
PMSA, JHK, and JR acknowledge financial support from the State Research Agency (AEI-MCINN) of the Spanish Ministry of Science and Innovation under the grant "The structure and evolution of galaxies and their central regions" with reference PID2019-105602GBI00/10.13039/501100011033, from the ACIISI, Consejer\'{i}a de Econom\'{i}a, Conocimiento y Empleo del Gobierno de Canarias and the European Regional Development Fund (ERDF) under grant with reference PROID2021010044, and from IAC project P/300724, financed by the Ministry of Science and Innovation, through the State Budget and by the Canary Islands Department of Economy, Knowledge and Employment, through the Regional Budget of the Autonomous Community. 
JR acknowledges funding from University of La Laguna through the Margarita Salas Program from the Spanish Ministry of Universities ref.~UNI/551/2021-May 26, and under the EU Next Generation. 
LVM acknowledges financial support from grants CEX2021-001131-S funded by MCIN/AEI/ 10.13039/501100011033, RTI2018-096228-B-C31 and PID2021-123930OB-C21 by MCIN/AEI/ 10.13039/501100011033, by “ERDF A way of making Europe” and by the "European Union" and from IAA4SKA (R18-RT-3082) funded by the Economic Transformation, Industry, Knowledge and Universities Council of the Regional Government of Andalusia and the European Regional Development Fund from the European Union. 
SC acknowledges funding from the State Research Agency (AEI-MCINN) of the Spanish Ministry of Science and Innovation under the grant “Thick discs, relics of the infancy of galaxies" with reference PID2020-113213GA-I00. MAF acknowledges support from FONDECYT iniciaci\'on project 11200107 and the Emergia program (EMERGIA20\_38888) from Consejería de Transformación Económica, Industria, Conocimiento y Universidades and University of Granada. 
PMSA and LVM acknowledge the Spanish Prototype of an SRC (SPSRC) service and support funded by the Spanish Ministry of Science, Innovation and Universities, by the Regional Government of Andalusia, by the European Regional Development Funds and by the European Union NextGenerationEU/PRTR. 
The SPSRC acknowledges financial support from the State Agency for Research of the Spanish MCIU through the "Center of Excellence Severo Ochoa" award to the Instituto de Astrofísica de Andalucía (SEV-2017-0709) and from the grant CEX2021-001131-S funded by MCIN/AEI/ 10.13039/501100011033. 
Based on observations made with the Isaac Newton Telescope operated on the island of La Palma by the Isaac Newton Group of Telescopes in the Spanish Observatorio del Roque de los Muchachos of the Instituto de Astrofísica de Canarias. The WFC imaging was obtained as part of the programs C163, C106, and C106/13B.
Based on observations collected at the European Organisation for Astronomical Research in the Southern Hemisphere under ESO programme(s) 098.B-0775(A), 093.B-0894(A).
Based on data collected at the Subaru Telescope and retrieved from the HSC data archive system, which is operated by Subaru Telescope and Astronomy Data Center at National Astronomical Observatory of Japan.

\end{acknowledgements}

\bibliographystyle{aa}

\onecolumn
\begin{appendix}

\section{SExtractor configuration parameters. }
\label{sec:annexe:Config}

The parameters that were not set to their default values are:\\

\raggedright{
\texttt{CHECKIMAGE\_TYPE SEGMENTATION \\
DEBLEND\_MINCONT  0.005 \\
DEBLEND\_NTHRESH 32 \\
BACK\_SIZE  20 \\
DETECT\_THRESH  0.9 \\ 
}}

\section{Images, masks, and surface brightness radial profiles.}
\label{sec:annexe:Images}

\begin{figure*}[!ht]
   \includegraphics[width=\size\textwidth]{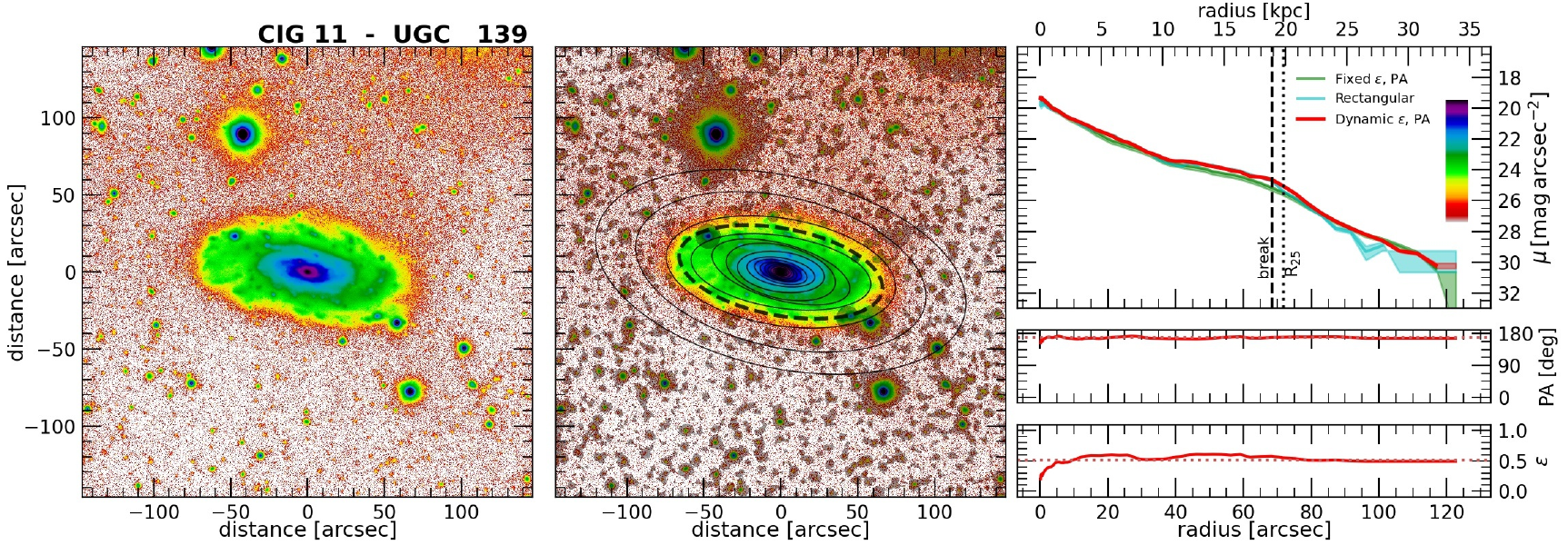}
    \includegraphics[width=\size\textwidth]{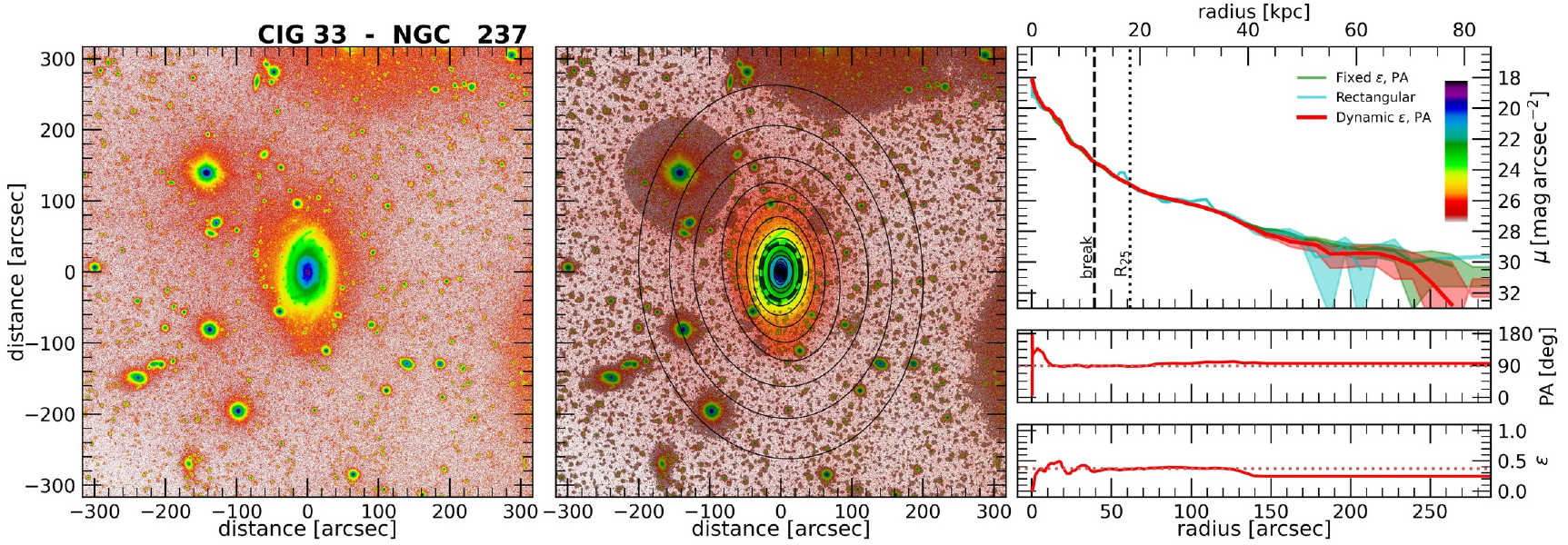}
    
    \caption{Left panel: r-band or luminance images for the galaxies in the sample. The ID of each galaxy is shown above. The scale of the colours represents the surface brightness of the image and the scale is shown in the profile panel (right) following the y-axis scale. Middle panel: Same image with the mask applied and the elliptical apertures of the dynamic positional angle and ellipticity profile. Right panel: Surface brightness profile (top), positional angle (middle), and ellipticity (bottom) as a function of the radius in arcseconds (bottom) and kpc (top). The red and green line the dynamic and fixed elliptical apertures, respectively. The blue line represents the profiles obtained from  rectangular apertures. The position of the disc break (if any) is shown with the vertical and elliptical aperture of the dashed line. We smooth CIG~613 image using the \software{FABADA} algorithm \citep[][]{fabada} to enhance the faint structures and we indicate with arrows the tidal stream found for this galaxy. More information about the figure can be found in Section~\ref{sec:Results}. }
\end{figure*}

\begin{figure*}[!ht]
    \includegraphics[width=\size\textwidth]{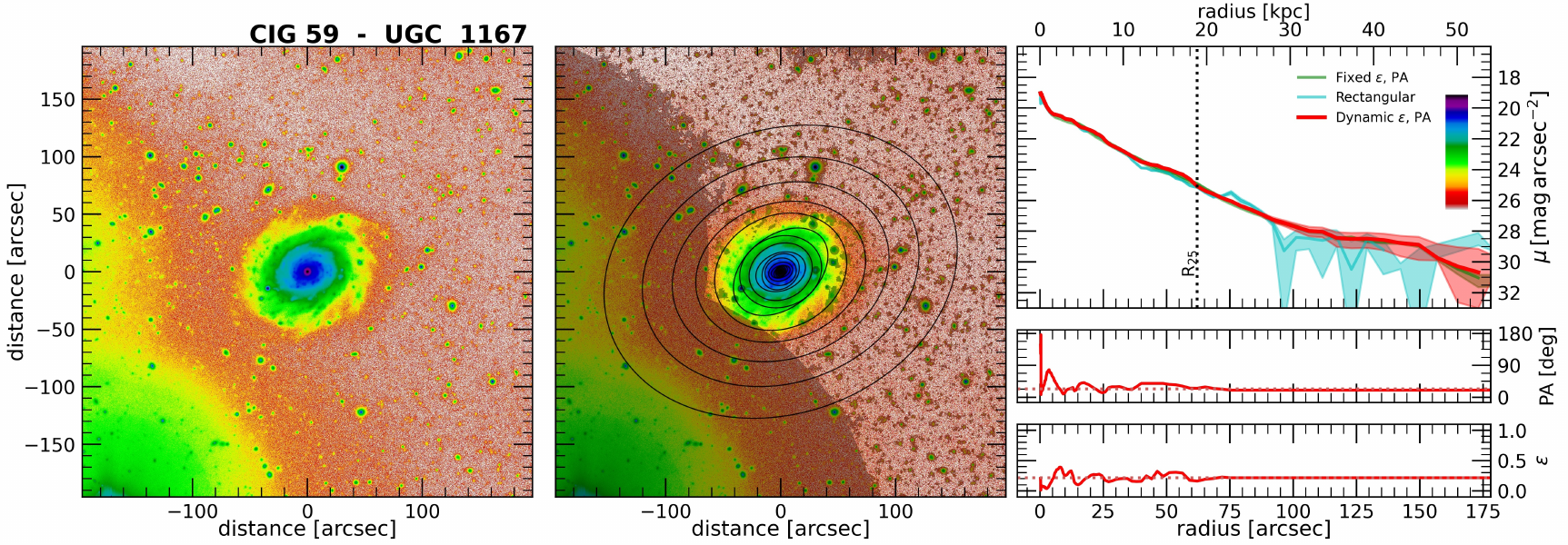}
    \includegraphics[width=\size\textwidth]{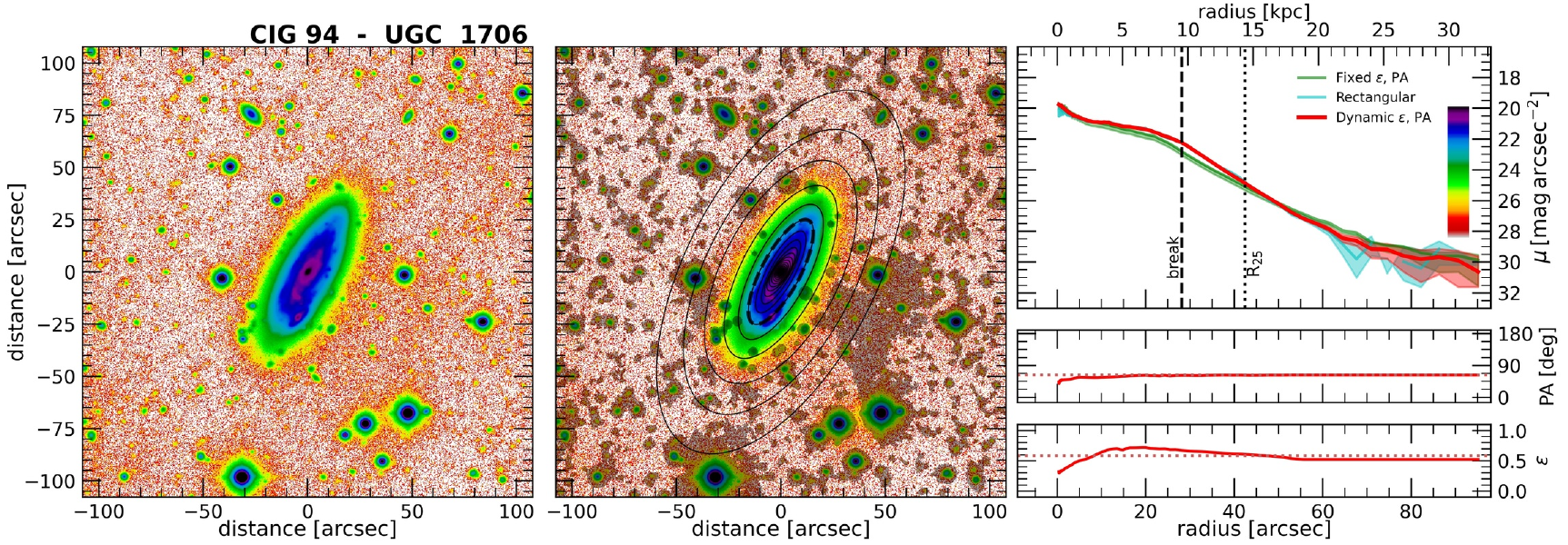}
    \includegraphics[width=\size\textwidth]{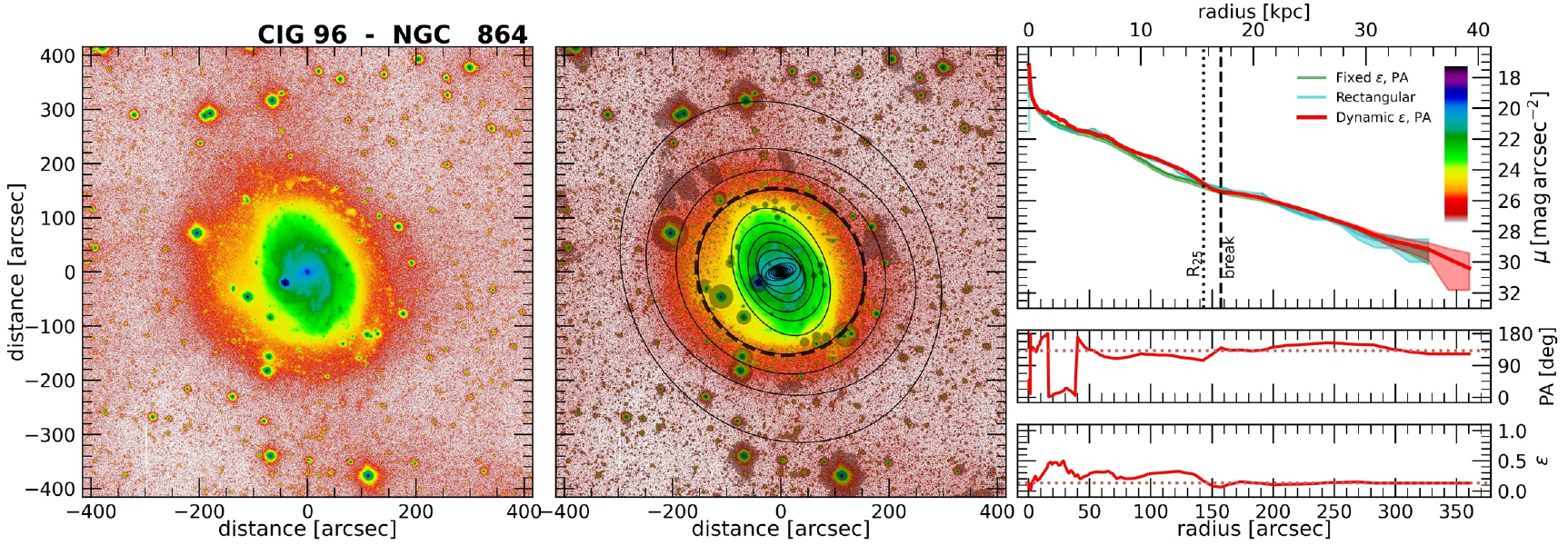}
    \includegraphics[width=\size\textwidth]{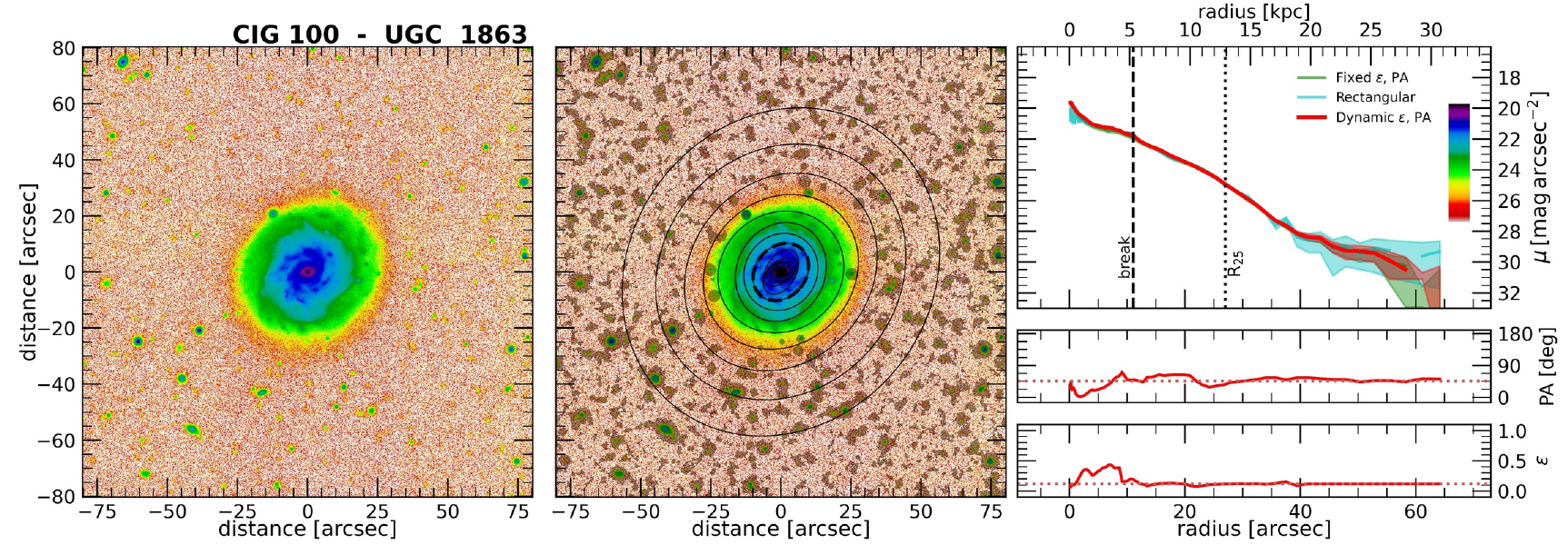}
   
\end{figure*}

\begin{figure*}[!ht]
     \includegraphics[width=\size\textwidth]{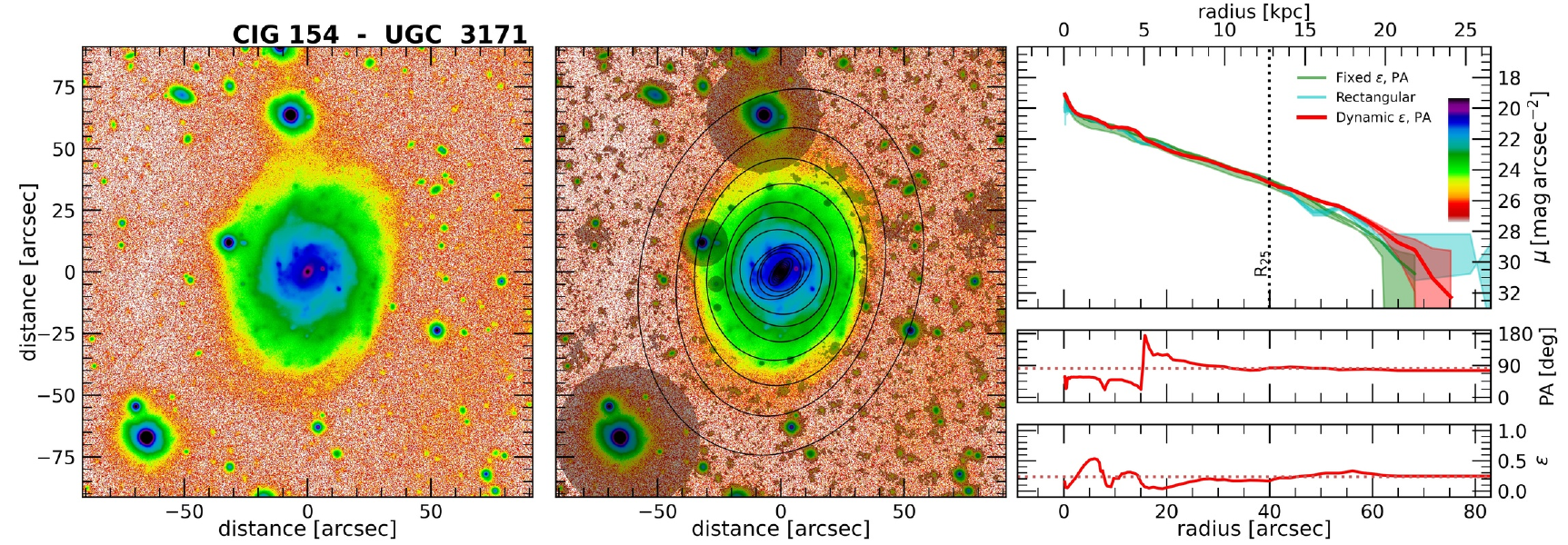}
     
    \includegraphics[width=\size\textwidth]{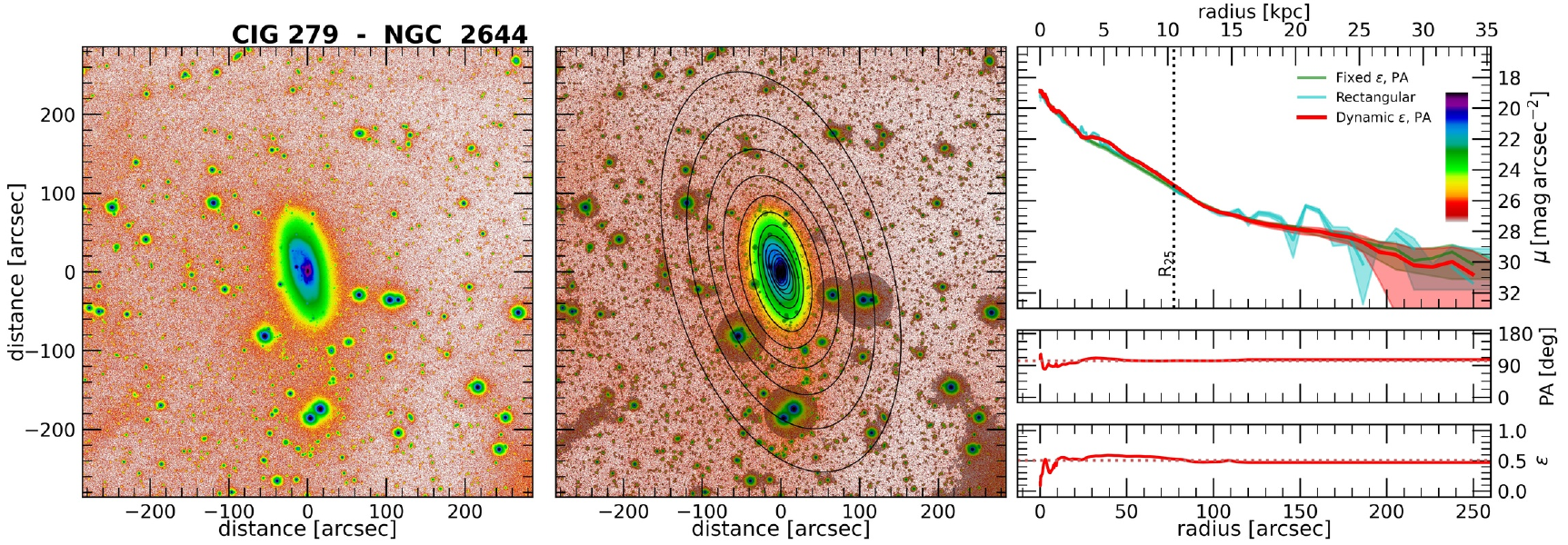}
    \includegraphics[width=\size\textwidth]{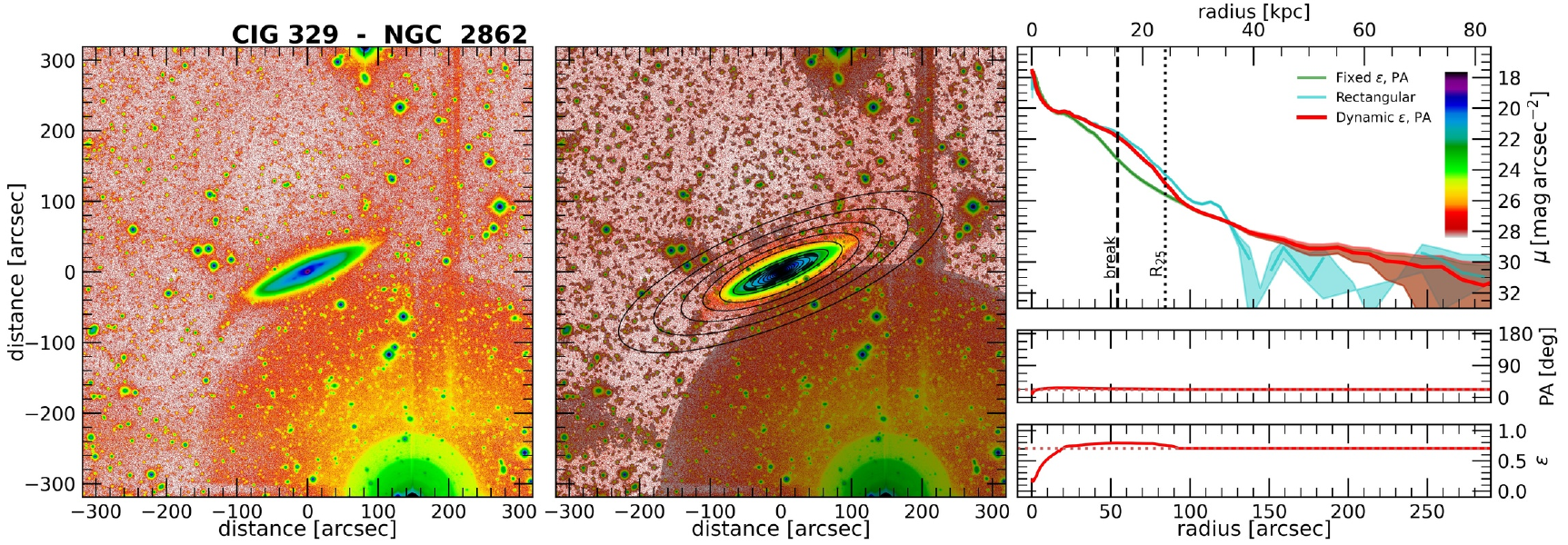}
    \includegraphics[width=\size\textwidth]{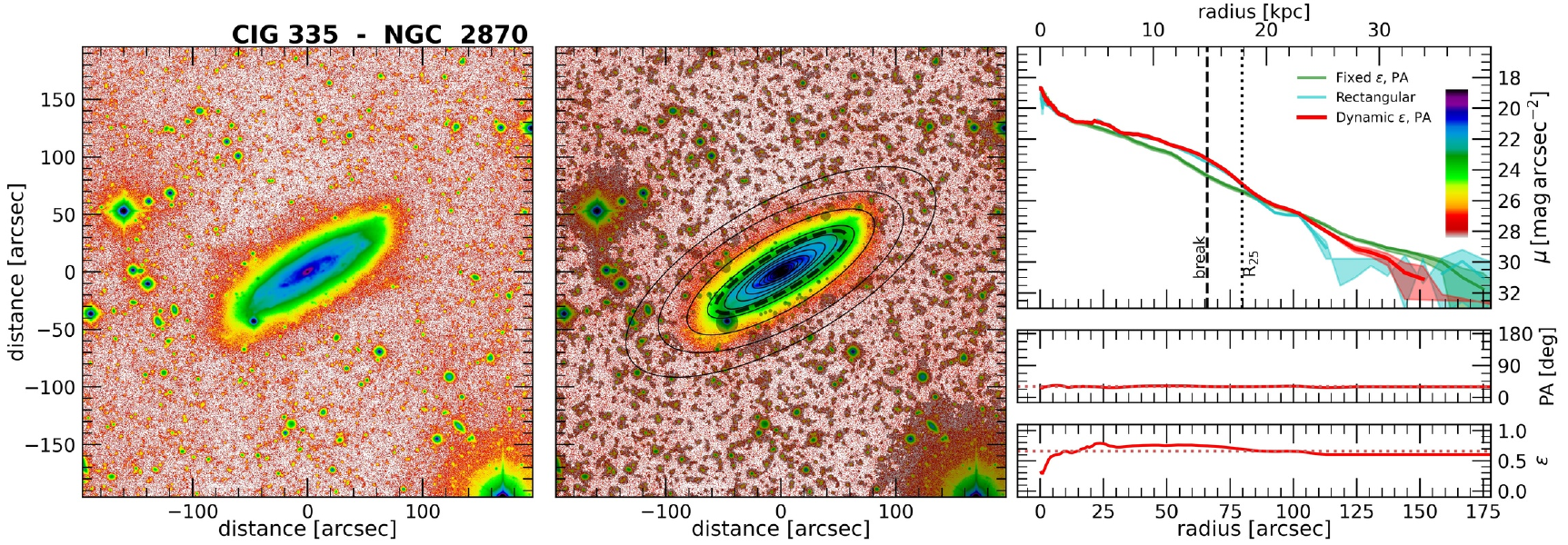}

\end{figure*}

\begin{figure*}[!ht]
    
    \includegraphics[width=\size\textwidth]{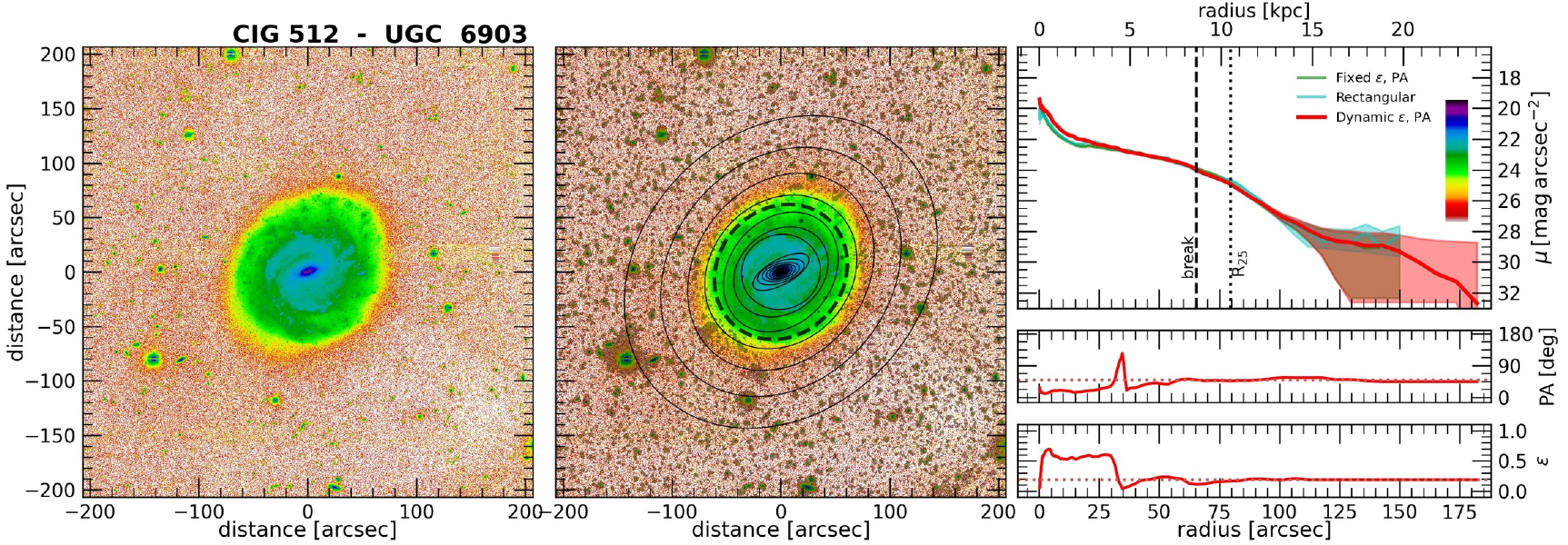}
    \includegraphics[width=\size\textwidth]{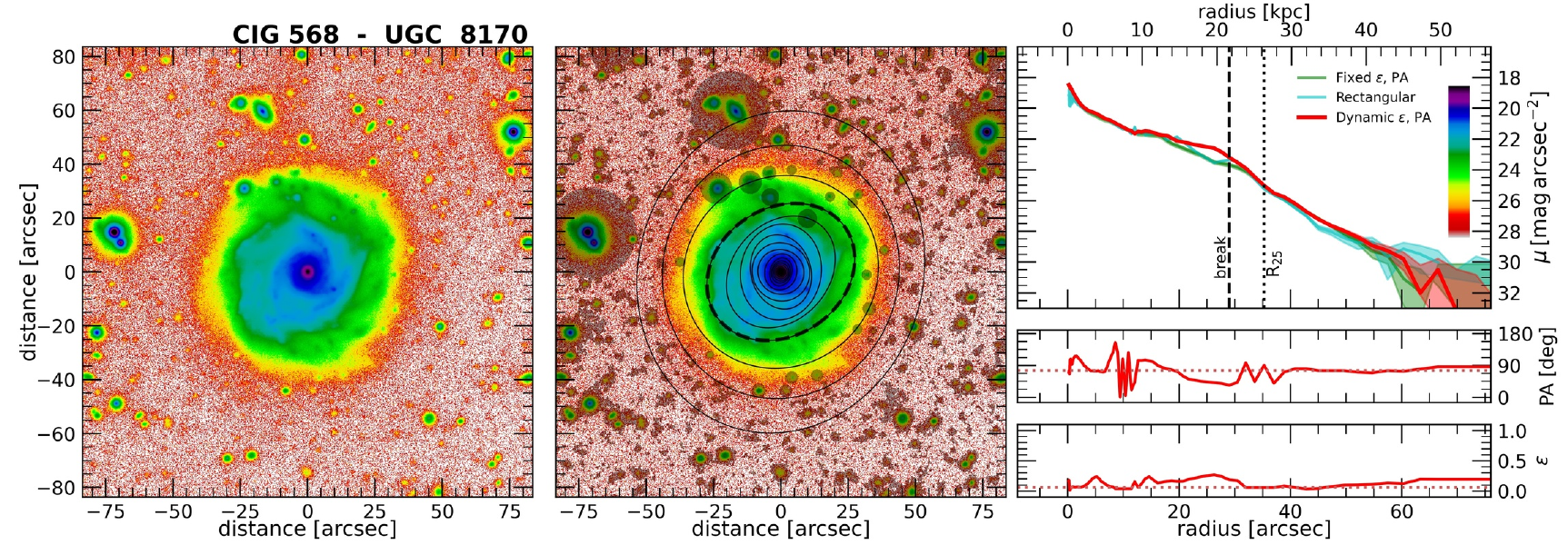}
    \includegraphics[width=\size\textwidth]{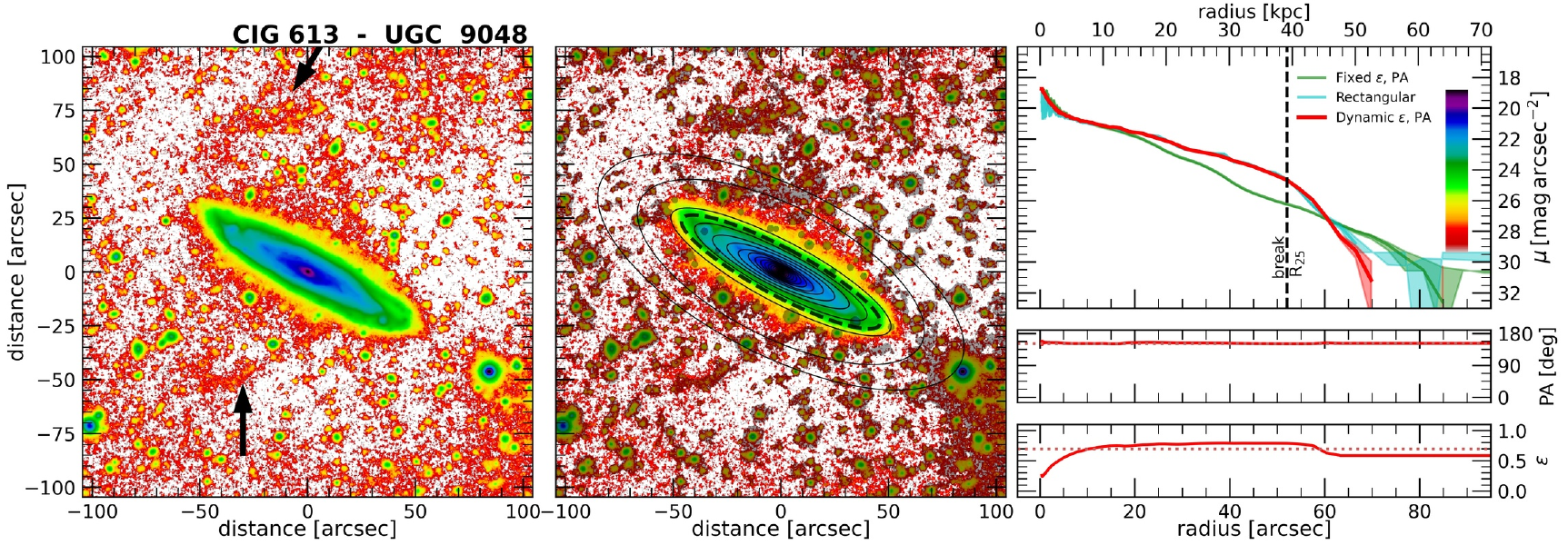}
    \includegraphics[width=\size\textwidth]{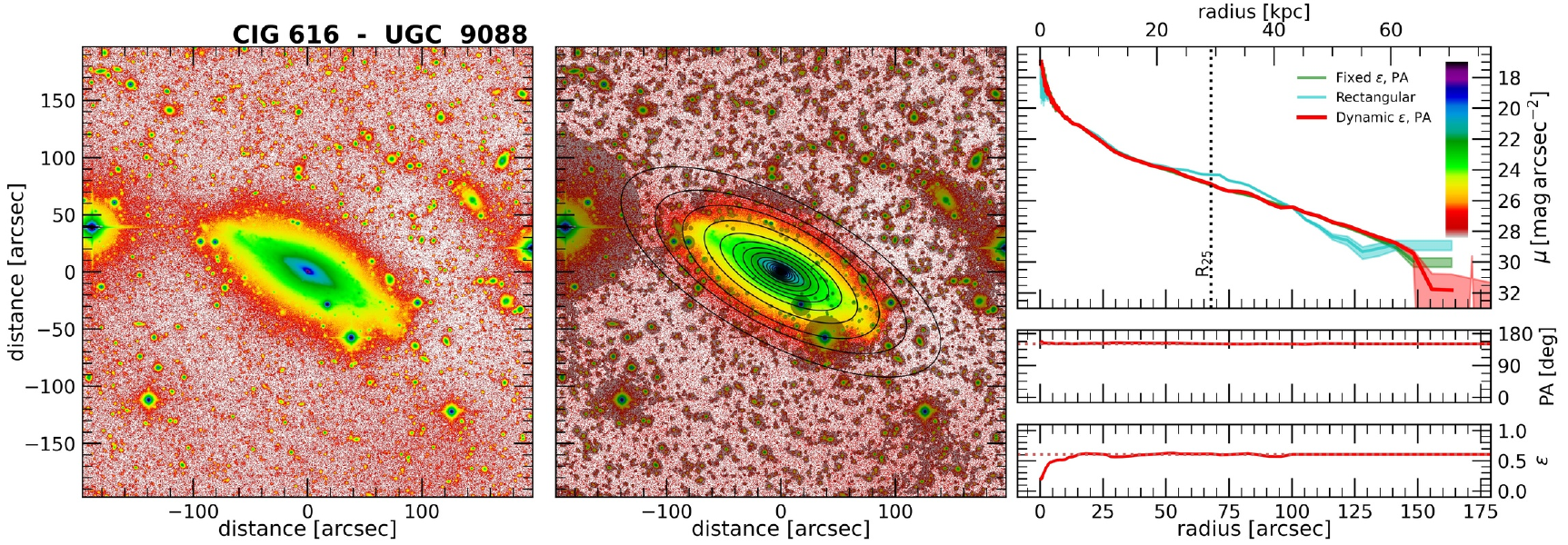}
\end{figure*}

\begin{figure*}[!ht]
    
    \includegraphics[width=\size\textwidth]{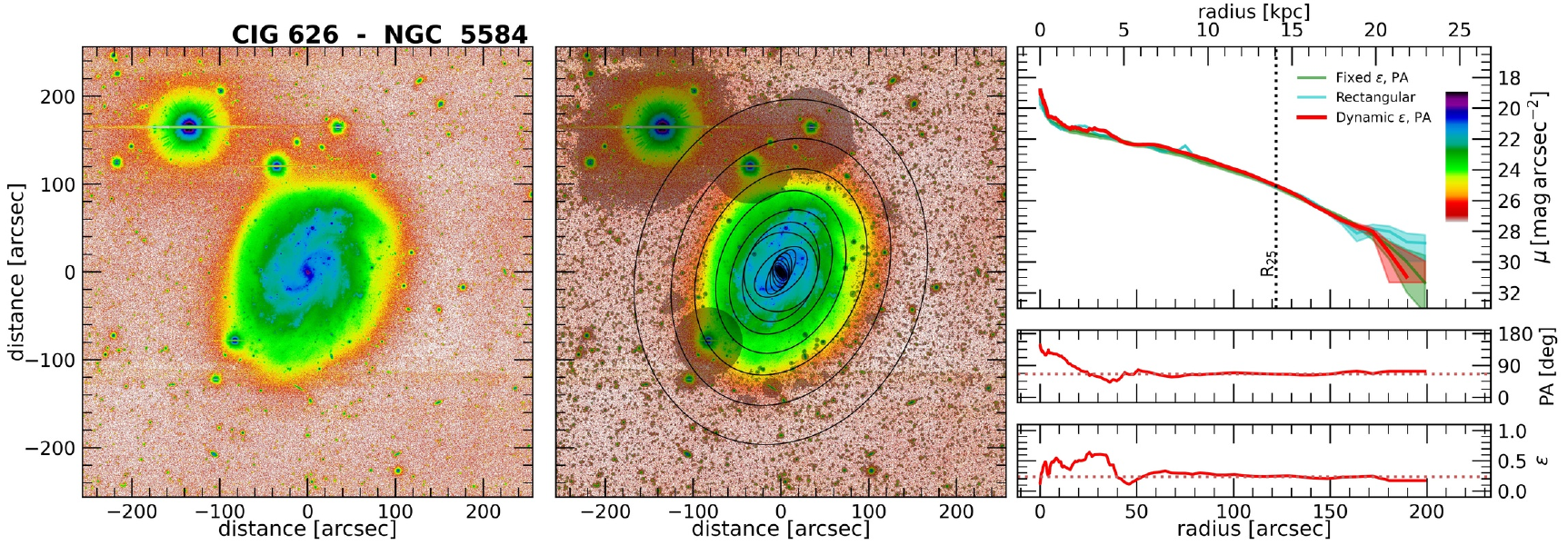}
    \includegraphics[width=\size\textwidth]{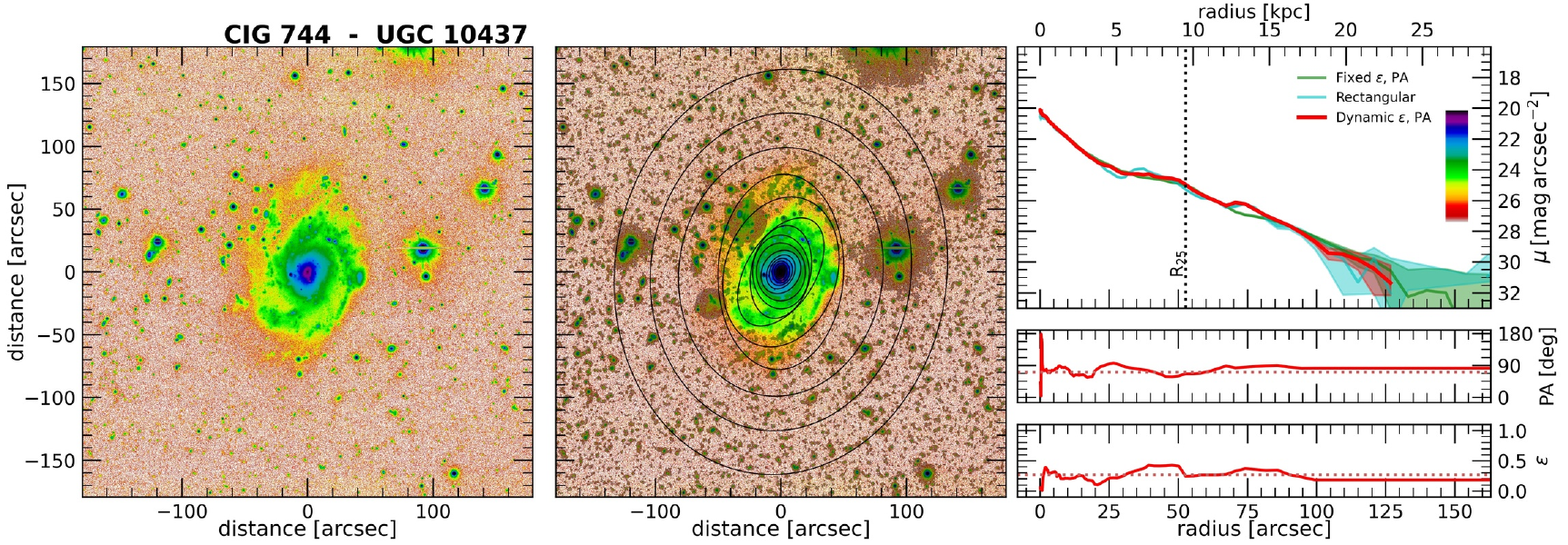}
    \includegraphics[width=\size\textwidth]{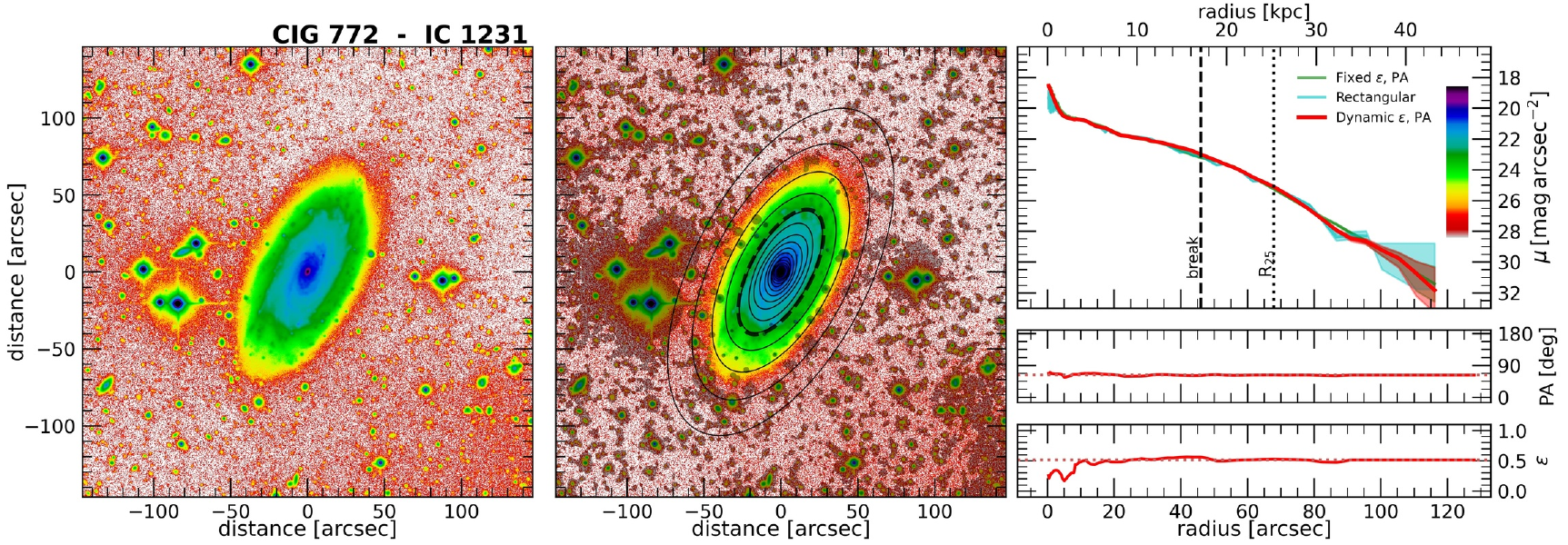}
   \includegraphics[width=\size\textwidth]{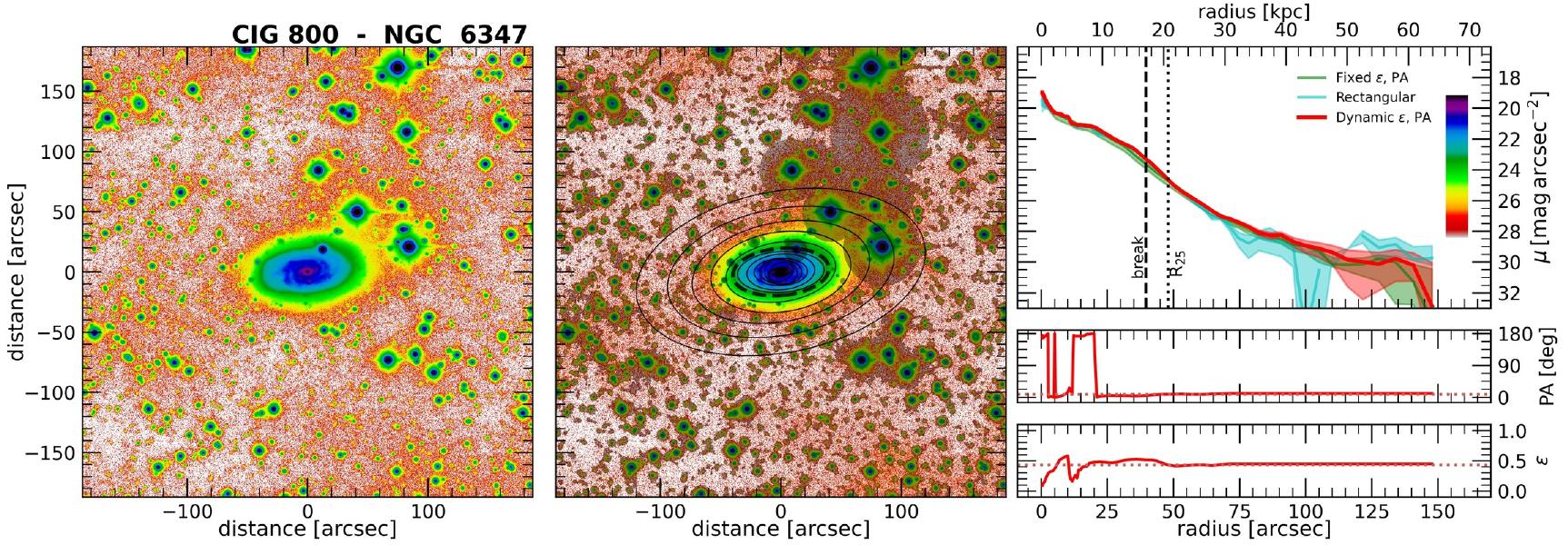}
\end{figure*}

\begin{figure*}[!ht]
    
    \includegraphics[width=\size\textwidth]{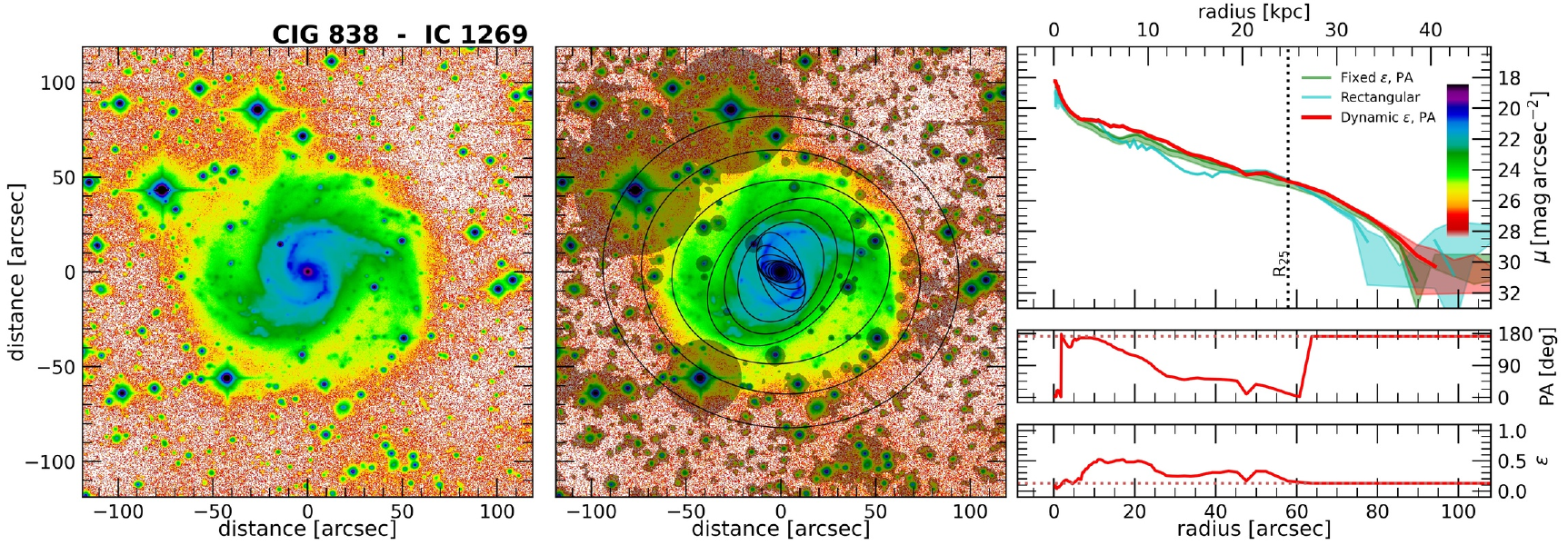}
    \includegraphics[width=\size\textwidth]{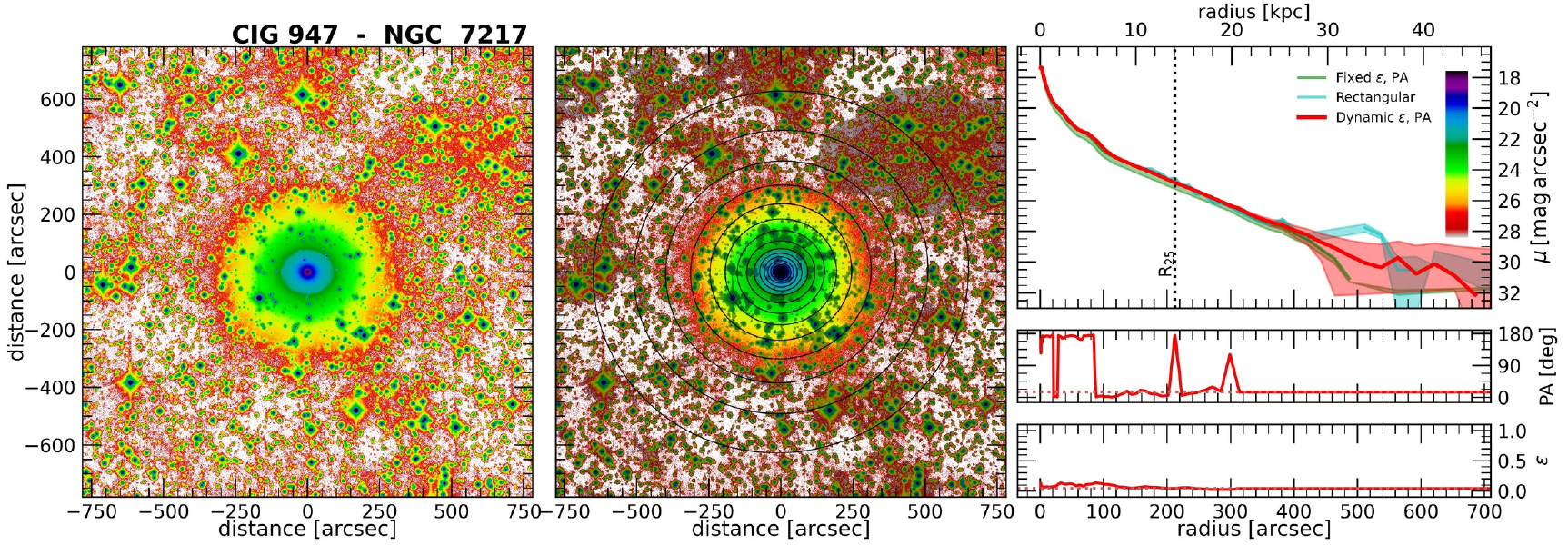}
    \includegraphics[width=\size\textwidth]{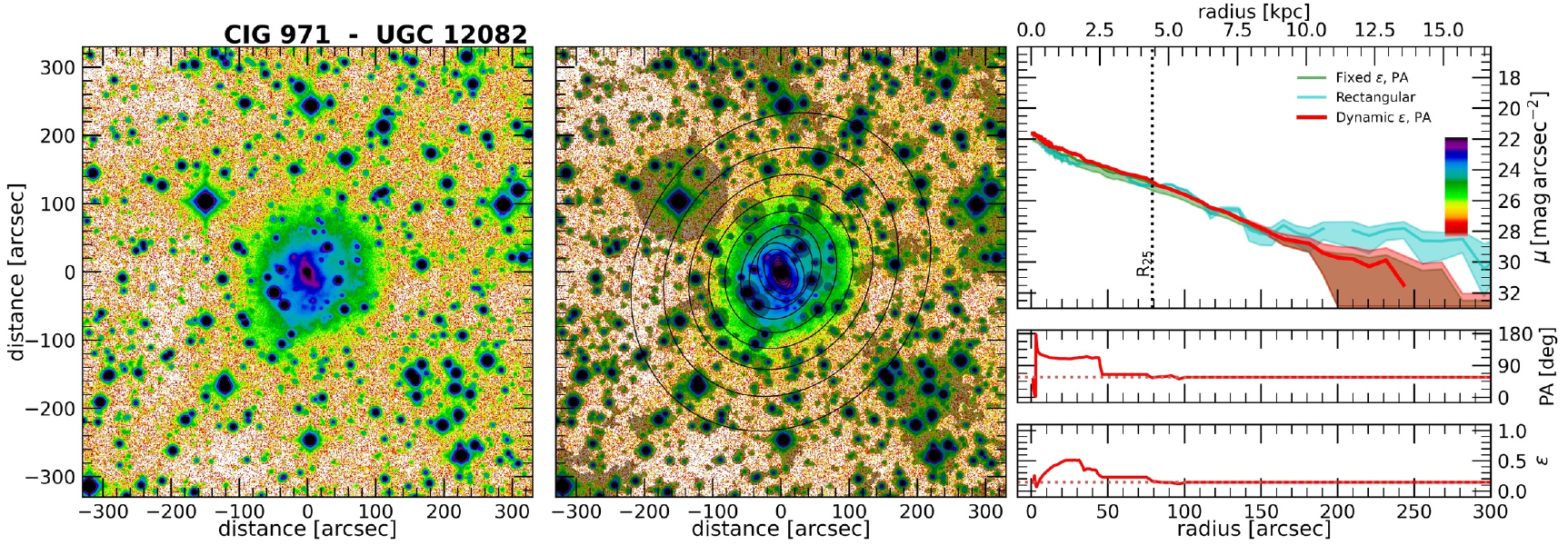}
    \includegraphics[width=\size\textwidth]{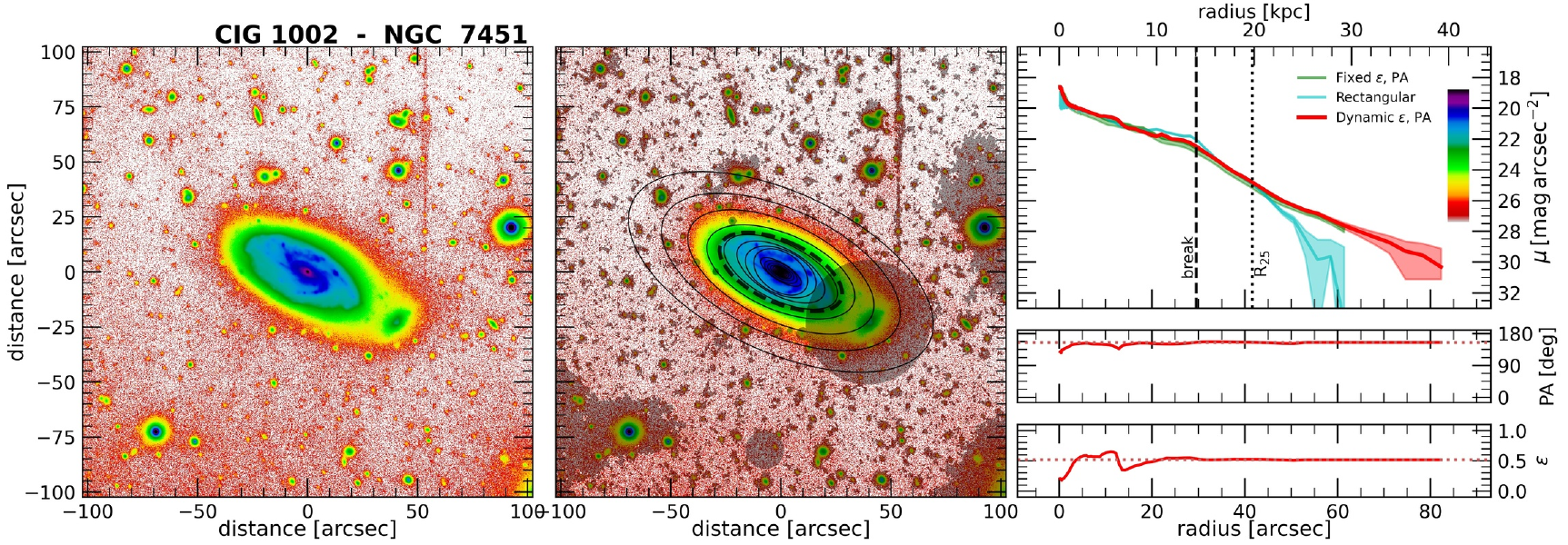}
\end{figure*}

\begin{figure*}[!ht]
    
    \includegraphics[width=\size\textwidth]{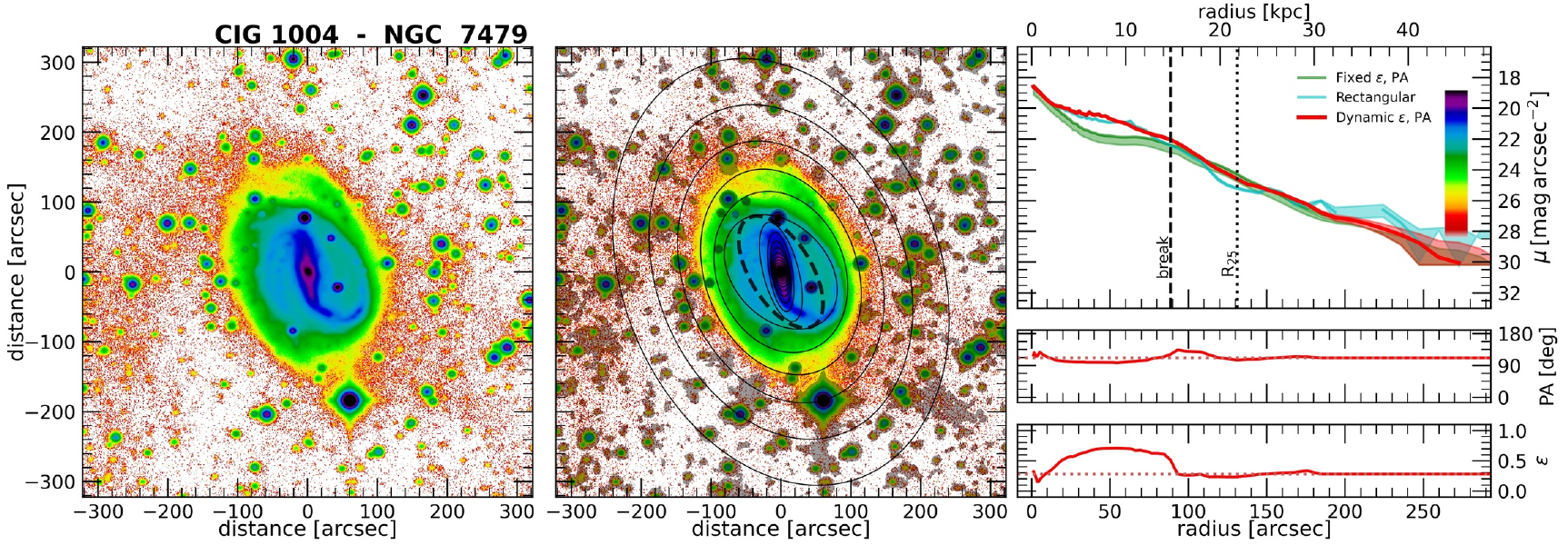}
    \includegraphics[width=\size\textwidth]{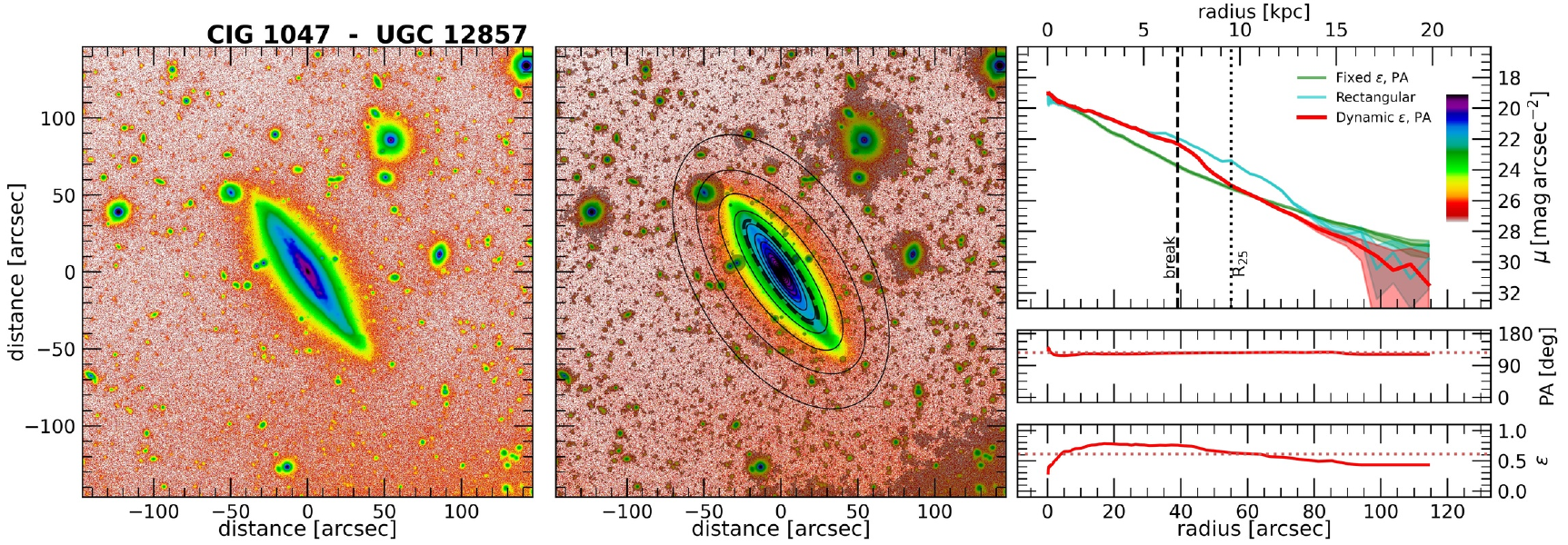}
\end{figure*}
\end{appendix}

\end{document}